\documentclass[a4paper,11pt,cleardoubleempty,footexclude,smallheadings]{scrreprt}

\pdfoutput=1

\usepackage[a4paper]{geometry}
\usepackage{cite}

\usepackage[utf8]{inputenc}
\let\oldbibliography\thebibliography
\renewcommand{\thebibliography}[1]{%
  \small\oldbibliography{#1}%
  \setlength{\itemsep}{0pt}%
}

\usepackage{amsmath,amssymb}
\usepackage{graphicx}
\usepackage[T1]{fontenc}
\usepackage{lmodern}

\newcommand{\longpage}[1][1]{}
\newcommand{\shortpage}[1][1]{}

\DeclareMathOperator{\Tr}{Tr}%
\DeclareMathOperator{\sign}{sign}%

\DeclareMathOperator{\diag}{diag}
\newcommand{\expp}[1]{ \mathop\mathit{e}\nolimits^{#1}}

\newcommand{\derf}[2]{\frac{\delta {#1}}{\delta {#2}}}

\newcommand{\av}[1]{\langle #1 \rangle}

\newcommand{\ud}[2][]{\textrm{d}^{#1}{#2}\,}

\newcommand{\vd}[2][]{\textrm{d}^{#1}{#2}}
\newcommand{\Eqref}[1]{Eq.~\eqref{#1}}
\newcommand{\ie}{\emph{i.e.}}
\newcommand{\eg}{\emph{e.g.}}
\renewcommand{\Re}{\mathop\mathrm{Re}\nolimits}
\renewcommand{\Im}{\mathop\mathrm{Im}\nolimits}
\newcommand{\vect}{\mathbf}

\newcommand{\udpi}[2][]{\frac{\textrm{d}^{#1}{#2}}{(2\pi)^{#1}}}

\newcommand{\GF}{G_\mathrm{F}}
\newcommand{\GD}{G_\mathrm{D}}
\newcommand{\SigmaR}{\Sigma_\mathrm R}

\newcommand{\Gret}{{G}_\mathrm R}

\newcommand{\GR}{\Gret}

\newcommand{\mph}{m_\mathrm{ph}}

\newcommand{\SigmaN}{\Sigma^{(1)}}
\newcommand{\ti}{{t_\text{i}}}
\newcommand{\tf}{{t_\text{f}}}

\newcommand{\GA}{G_\text{A}}
\newcommand{\pplus}{{\scriptscriptstyle (+)}}
\newcommand{\pminus}{{\scriptscriptstyle (-)}}
\newcommand{\meff}{m_\text{eff}}
\newcommand{\phir}{\bar\phi}
\newcommand{\pir}{\bar\pi}
\newcommand{\ar}{\bar{a}}

\allowdisplaybreaks[3]

\providecommand{\abs}[1]{\lvert#1\rvert}

\newcommand{\fr}{^{(0)}}

\providecommand{\citep}{\cite}

\begin{document}

\begin{titlepage}
	\linespread{1.05}
	\centering
	\sffamily
	\mbox{}
	\vfill
	{\Large \bfseries Quasiparticle excitations in relativistic quantum field theory\par}
	\vspace{2em}
	{\large\rmfamily  Daniel Arteaga\par}
	\vspace{2em}
	{\rmfamily \itshape \normalsize
Departament de F\'\i sica Fonamental and Institut de Ci\`encies del Cosmos, \\ Facultat de F\'\i sica, Universitat de Barcelona, \\ Av.~Diagonal 647, 08028 Barcelona (Spain)\par	}
	\vspace{1em}
	\rmfamily e-mail: \texttt{darteaga@ub.edu}
	\vspace{3em}
	\begin{center}
		\small \textbf{Abstract}
	\end{center}
	\begin{quotation}
		\rmfamily \normalsize We analyze the particle-like excitations arising in relativistic field theories in states different than the vacuum. The basic properties characterizing the quasiparticle propagation are studied using two different complementary methods. First we introduce a frequency-based approach, wherein the quasiparticle properties are deduced from the spectral analysis of the two-point propagators.
		Second, we put forward a real-time approach, wherein the quantum state corresponding to the quasiparticle excitation is explicitly constructed, and the time-evolution is followed. 
 		Both methods lead to the same result: the energy and decay rate of the quasiparticles are determined by the real and imaginary parts of the retarded self-energy respectively.  Both approaches are compared, on the one hand, with the standard field-theoretic analysis of particles in the vacuum and, on the other hand, with the  mean-field-based techniques in general backgrounds. 
	\end{quotation}
	\vfill \vfill
	\mbox{}
\end{titlepage}



%

\tableofcontents

\chapter{Introduction}

In this paper we examine the elementary particle-like excitations in generic quantum states, as seen from the viewpoint of relativistic quantum field theory. We will refer to those excitations as \emph{quasiparticles}. Quasiparticles are one of the most ubiquitous concepts in physics, appearing in many different contexts with slightly different meanings, for instance in Bose-Einstein condensation \cite{
Dalfovo99}, quantum liquids \cite{PinesNozieres89}, superconductivity \cite{KaplanEtAl76}, 
and, more in general, in condensed matter field theory \cite{Abrikosov,AltlandSimons}, thermal field theory \cite{LeBellac,Kapusta} and $N$-body quantum mechanics (non-relativistic field theory) \cite{FetterWalecka}. We shall not by any means attempt to review the concept of quasiparticle in this paper. Instead, we shall limit ourselves to studying the dynamics of particle-like excitations arising in relativistic scalar field theories in states different than the vacuum, focusing on those properties which can be extracted from  the two-point correlation functions. 

In order to analyze the particle or quasiparticle properties from a second-quantized perspective the standard procedure is to add an elementary particle-like excitation to the system, and to analyze the subsequent dynamical evolution of the field mode corresponding to the initial particle momentum. With non-relativistic excitations this procedure is clear, given that the particle number is preserved along the evolution and the particle concept is well-defined for all times. 
In this context it can be readily obtained that the energy and decay rate  of the (quasi)particle excitations can be recovered from the real and imaginary parts of the poles of the two-point propagators respectively \cite{FetterWalecka}. 

With relativistic excitations the procedure is less direct since the number of relativistic particles in a given field state fluctuates. Properly speaking, in presence of interactions particles are only well-defined in the asymptotic limit, where the interaction between them can be neglected. Single particle excitations in the vacuum constitute a particular case, since Lorentz symmetry is enough to fully characterize their properties \cite{WeinbergQFT}. It is then a textbook result to show that in the vacuum the energy of a single particle particle is also given by the location of the pole of the Feynman propagator, similarly to the non-relativistic case. For unstable particles the treatment is less rigorous since properly speaking no asymptotic states can be associated to them. Nevertheless by analogy with the stable case, and appealing to the optical theorem, it can be additionally argued that the imaginary part of the pole of the propagator determines the lifetime of the particle.

In non-vacuum states, where it is not possible to bring up symmetry considerations, it is  a priori not completely obvious how the relativistic quasiparticles should be treated within a second-quantized formalism. One possibility is to simply forget about quasiparticles and to study the dynamics of the mean field under arbitrary small perturbations, using the linear response theory \cite{Kapusta,LeBellac,FetterWalecka,Reichl}. The response of the mean field is characterized by a frequency and a decay rate, which are respectively connected to the real and imaginary parts of the retarded propagator. This constitutes the standard way  of analyzing relativistic excitations in non-vacuum states \cite{LeBellac,FetterWalecka,Weldon99}, although for typical particle-like excitations the expectation value of the field vanishes, and therefore the excitations considered by the linear response method do not correspond to elementary quasiparticles. In any case, the approach based on the linear response theory is appropriate to analyze another different regime, the hydrodynamic or fluid regime \cite{Forster,Groot,SonStarinets07}.

Whenever single quasiparticle excitations are important, a second possibility is trying to develop an appropriate quasiparticle framework, following what is done either in non-relativistic field theory or in relativistic field theory in the vacuum. This will be the main object of analysis of this paper. We will investigate several questions: first, which is the field state corresponding to quasiparticle excitations; second, whether quasiparticle excitations appear naturally in physical situations; third, how to extract their properties from the field-theoretic correlation functions, and, finally,  what is the range of applicability of the quasiparticle formalism. We will however not discuss under which conditions a given field theory in a given field state possesses quasiparticle excitations: it will be simply assumed that the field interaction and state are such that a quasiparticle regime exists.

There have been some partial results in this direction. Weldon \cite{Weldon83} recognized that in thermal field theory the imaginary part of the pole of the retarded propagator is linked to the decay rate of the elementary excitations. Donoghue \emph{et al.} \cite{DonoghueEtAl85} discussed the role of the the real part of the pole of the propagator as related to the energy of the quasiparticles, and the fact that it is not necessarily Lorentz-invariant. Narnhofer \emph{et al.}\ \cite{NarnhoferEtAl83} studied the properties of the quasiparticles in thermal states.   Weldon \cite{Weldon98} and Chu and Umezawa \cite{ChuUmezawa93} attempted a reformulation of thermal field theories in terms of what they called stable quasiparticles, the latter using a thermo-field dynamics formalism.
Also in the thermo-field dynamics context, Nakawaki \emph{et al.}\ \cite{NakawakiEtAl89} studied quasiparticle collisions (a subject which will not be discussed in this paper). Finally, let us mention that Greiner and Leupold \cite{GreinerLeupold98}, among other things, studied the conditions for the existence of a quasiparticle regime and analyzed the fluctuations of the  quasiparticle number.

The kinetic approach to field theory by Calzetta and Hu \cite{CalzettaHu88} also goes beyond the mean-field treatment. Calzetta and Hu focus on the dynamics of the propagator by studying the two-particle irreducible  effective action, which leads to equations of motion for the correlation functions, from where the equations of motion for the quasiparticle occupation numbers are extracted, thereby connecting the second-quantized formalism with a first-quantized statistical description. In this paper we will strictly limit to a second-quantized approach; see Ref.~\cite{ArteagaThesis} for some comments on the connection with the first-quantized analysis.



It should be noted that, since we shall work within a second-quantized formalism, we will not study the quasiparticle themselves but rather the field modes corresponding to the initial momentum of the quasiparticles. We will therefore not attempt to follow the quasiparticle trajectory. It will be important to bear in mind that the appearance of a dissipative term does not necessarily mean a quasiparticle decay, but most of the times it simply indicates a change of momentum of the quasiparticle. 

The paper is organized as follows. In the remaining of this section we introduce some of the tools and concepts that will be  used in the paper. Of particular interest is the analysis of the two-point propagators in general backgrounds.  In Sec.~\ref{II} we recall how the particle interpretation can be recovered in field theory in the vacuum, introducing the two techniques that later on will be employed for the analysis of the quasiparticles. In Sec.~\ref{III} we develop a spectral approach to the analysis of the quasiparticle excitations in field theory, in parallel to the procedure in the vacuum. In Sec.~\ref{IV} we present a complementary real-time analysis of the same problem, namely we study the time evolution of the relevant observables in the presence of quasiparticle excitations. In the process we discuss the form of the quantum states associated to quasiparticles, and analyze the appearance of these states in physical situations. In Sec.~\ref{V} we recall the standard mean-field-based techniques and compare them with our methods and results. Finally, in Sec.~\ref{VI} we summarize and discuss the main points of the paper. Appendices contain background reference material as well as technical details of some calculations.

Throughout the paper we use a metric with signature $(-,+,+,+)$, work with a system of natural units with $\hbar =c =1$, denote quantum mechanical operators with a hat, and use a volume-dependent normalization in the definition of the field modes [see \Eqref{ModeDecomp} below]. The same symbol will be later used for a quantity and its Fourier transform whenever there is no danger of confusion.

\section{Quasiparticles in relativistic field theory}\label{sect:QuasiDef}


\index{Quasiparticle|(}
 For the purposes of this paper we consider that a quasiparticle  is a particle-like excitation which travels in some background and which is characterized by the following properties:
\begin{enumerate}
	\item It has some characteristic initial energy $E$. The fluctuations of the energy are much smaller than this characteristic value: $\Delta E \ll |E|$.
	\item It has some characteristic initial momentum $\vect p$. The  fluctuations of the momentum are much smaller than the characteristic energy:  $|\Delta \vect p| \ll |E|$.
	\item It has approximately constant energy and momentum during a long period of time $T=1/\Gamma$, before it starts to decay. Here ``long'' means that the decay rate $\Gamma$ has to be much smaller than the de Broglie frequency of the quasiparticle: $\Gamma \ll |E|$. 
	\item It is elementary, meaning that it cannot be decomposed in the (coherent or incoherent) superposition of two or more entities, having each one separately the same three properties above.
\end{enumerate}
Notice that the third property is somewhat redundant, since by the time-energy uncertainty principle the energy fluctuations are at least given by the decay rate: $\Delta E \gtrsim \Gamma$.
The quasiparticle is essentially characterized by the energy $E$, the momentum $\vect p$ and the decay rate $\Gamma$. Besides that, the quasiparticle can be characterized by other quantum numbers such as the spin (although in this paper we shall only deal with scalar quasiparticles).


If the initial background state has large momentum or energy fluctuations the perturbed state inherits them, and therefore the requirement that the fluctuations of the momentum and energy of the quasiparticles are small cannot be fulfilled. For thermal and, more generally, for Gaussian states, we will see that momentum fluctuations are comparable to the average momentum when the occupation numbers are of order one. Therefore for bosonic systems the quasiparticle description of Gaussian states requires relatively small occupation numbers.
When the occupation numbers are of order one or larger, a quasiparticle description might not be very adequate and it might be more useful to move to a hydrodynamic  description \cite{Groot,Forster,SonStarinets07} (see Sec.~\ref{VI} for a further analysis of this point). 

\index{Collective mode}
\index{Quasiparticle!in the strict sense}
In any case, quasiparticle excitations may exist even in strongly correlated systems. The excitations in these systems are usually radically different to the original particles which constitute the medium. For instance, the spin and mass of the quasiparticles might have nothing to do with the original vacuum particles. The quasiparticles which bear no resemblance with the original particles constituting the medium, and which usually involve the system as a whole, are also called \emph{collective modes} \cite{FetterWalecka}. Sometimes in the same system  the collective modes can coexist with the quasiparticles directly corresponding to the medium constituents.

In this paper we will consider the quasiparticle excitations of a single scalar field $\phi$ in a given background state.  This field can have two different interpretations depending on the context. First, the field $\phi$ can have a straightforward interpretation as the fundamental field whose excitations in the vacuum give rise to the particles, and whose excitations in states different than the vacuum constitute the corresponding quasiparticles. For instance, this would be an appropriate interpretation when studying the behavior of a pion in a thermal bath. But the field $\phi$ can have a second different interpretation as an effective field whose excitations correspond to the quasiparticles propagating in the system, not necessarily having any direct correspondence with the fundamental fields. This second interpretation is more general and can accommodate for the collective modes. For instance, the field $\phi$ could represent the field of sound wave excitations in a Bose-Einstein condensate. In any case, in this paper we will consider the scalar field $\phi$ as given and will not investigate its relation with the fundamental constituents of the system.

\section{Dispersion relations}

The \emph{dispersion relation} is the expression of the energy of the quasiparticle as a function of the momentum, namely
\begin{subequations}
\begin{equation}
	E = E_\vect p,
\end{equation}
where $E$ is the energy of the quasiparticle excitation, and $\vect p$ is the momentum, with the assumption of small spreads. 
The \emph{effective mass} is the value of the energy at zero momentum,
$
	\meff = E_{\vect p= \vect 0}
$. When the states are thermal, the effective mass is called the \emph{thermal mass}.
For a unstable system with energy $E$ and decay rate $\Gamma$, we define the complex \emph{generalized energy} $\mathcal E$ as
$
	\mathcal E^2 := E^2 - i E \Gamma
$. 
The \emph{generalized dispersion relation} is the function giving the generalized energy of a quasiparticle in terms of the momentum: 
\begin{equation} \label{GenDispRel}
	\mathcal E^2 = E_\vect p^2 - i E_\vect p \Gamma_\vect p.
\end{equation}
\end{subequations}
Notice that the imaginary part of the generalized dispersion relation places a lower bound on the uncertainty of the real part.

In flat spacetime in the vacuum, the propagation of a stable particle is fully characterized by   the physical mass $\mph$:
\begin{subequations} \label{VacuumDispRel}
\begin{equation}
	E^2 = \mph^2 + \vect p^2.
\end{equation}
If the particle is unstable, the generalized dispersion relation is determined by the physical mass and the decay rate in the particle rest frame  $\gamma$:
\begin{equation}
	\mathcal E^2 = \mph^2 + \vect p^2 - i \mph \gamma.
\end{equation}
\end{subequations}
The decay rate must be much smaller than the particle mass; otherwise one would speak of resonances rather than unstable particles. In any case, the generalized dispersion relation can be extracted from the location of the poles in the momentum representation of the Feynman propagator. In other words, the Feynman propagator can be approximated around the particle energy as: 
\begin{equation} \label{IntrodFeynman}
	\GF(\omega,\vect p) \approx \frac{-iZ}{-\omega^2 + \vect p^2 + \mph^2 -  i \mph \gamma}, \quad \text{for } \omega^2 \sim \vect p^2 + \mph^2.
\end{equation} 

Several differences with respect to the vacuum case arise when studying the propagation in a non-vacuum state. First, in general the dispersion relation needs not be of any of the forms \eqref{VacuumDispRel}, and therefore the effective mass does not determine by itself the dispersion relation. The background constitutes a preferred reference frame, thus breaking the global Lorentz invariance, and, because of the interaction with the environment modes, the quasiparticle energy is affected by the background in a way such that the dispersion relation is no longer Lorentz invariant ~\cite{Ojima73,ArteagaParentaniVerdaguer04a,ArteagaParentaniVerdaguer05,Requardt08}.
Moreover, even if  some of the basic features of the particle propagation can be encoded in the form of a dispersion relation, one should bear in mind that a much richer phenomenology appears in general (including, for instance, scattering, diffusion and decoherence). Additionally, let us point out that it is not completely obvious how in general the dispersion relations should be extracted from the poles of a propagator, in a similar way to \Eqref{IntrodFeynman}; we will also address this point in this paper.

\section{Propagators and self-energies in an arbitrary background state} \label{sect:GSigmaGeneral}


In the vacuum the analysis of the Feynman propagator is usually sufficient. In a generic state $\hat\rho$ this is not usually the case, and an analysis of the different propagators is in order \cite{ChouEtAl85,Lawrie88,DasThermal}. The Feynman propagator, positive and negative Whightman functions and Dyson propagator,
\index{Propagator!Feynman}
\index{Propagator!Whightman}
\index{Propagator!Dyson}
\begin{subequations} \label{CorrFunct}
\begin{align}
    G_{11}(x,x') &= G_\text{F}(x,x') := \Tr{\big(
\hat\rho\, T \hat\phi(x) \hat\phi(x') \big)
}, \\
    G_{12}(x,x') &= G_{+}(x,x') := \Tr{\big(
\hat\rho\,  \hat\phi(x) \hat\phi(x') \big)
}, \\
    G_{21}(x,x') &=G_{-}(x,x') := \Tr{\big(
\hat\rho\,  \hat\phi(x') \hat\phi(x) \big)
}, \\
    G_{22}(x,x') &=G_\text{D}(x,x') := \Tr{\big(
\hat\rho\,  \widetilde T \hat\phi(x) \hat\phi(x')\big)
 },
\end{align}
\end{subequations}
appear in the closed time path (CTP) formalism  (which is natural when dealing with non-vacuum states; see appendix \ref{app:CTP}), and can be conveniently organized in a $2\times 2$ matrix $G_{ab}$, the so-called \emph{direct basis}:
\begin{equation}
	G_{ab}(x,x') = 
		\begin{pmatrix}
			\GF(x,x') & G_-(x,x') \\
			G_+(x,x') & G_\text{D}(x,x')
 		\end{pmatrix}.
\end{equation}

We may also consider the Pauli-Jordan or commutator propagator,
\index{Propagator!Pauli-Jordan}
\begin{subequations}\label{CorrFunct2}
\begin{equation}
 G(x,x') := \Tr{\big(
\hat\rho\,[  \hat\phi(x) ,\hat\phi(x')] \big)
},
\end{equation}
and the Hadamard or anticonmutator function
\index{Propagator!Hadamard}
\begin{equation}
	G^{(1)}(x,x') := \Tr{\big(
\hat\rho\,\{  \hat\phi(x) ,\hat\phi(x') \} \big)}.
\end{equation}
\end{subequations}
For linear systems\footnote{By linear systems we mean systems whose Heisenberg equations of motion are linear. These correspond either to non-interacting systems or to linearly coupled systems.} the Pauli-Jordan propagator is independent of the state and carries information about the system dynamics. Finally, one can also consider the retarded and advanced propagators,
\index{Propagator!retarded}
\index{Propagator!advanced}
\begin{subequations}\label{CorrFunct3}
\begin{align}
    G_\mathrm {R}(x,x') &= \theta(x^0-x'^0) G(x,x') =\theta(x^0-x'^0) \Tr{\big(
\hat\rho\, [\hat \phi(x),\hat\phi(x')]\big)}, \label{RetardedProp}\\
    G_\mathrm {A}(x,x') &= \theta(x'^0-x^0) G(x,x') =\theta(x'^0-x^0) \Tr{\big(
\hat\rho\, [\hat \phi(x),\hat\phi(x')]\big)},
\end{align}
\end{subequations}
which also do not depend on the the state for linear systems.

\index{Keldysh basis}
\index{Physical basis|see{Keldysh basis}}
By doing a unitary transformation of the propagator matrix it is possible to work in the so-called \emph{physical} or \emph{Keldysh basis} \cite{DasThermal,EijckKobesWeert94},
\begin{equation}\label{Keldysh}
	G'_{a'b'}(x,x') = 
		\begin{pmatrix}
			G^{(1)}(x,x') & G_\text R(x,x') \\
			G_\text A(x,x') & 0
 		\end{pmatrix},
\end{equation}
 For linear systems, the off-diagonal components of the Keldysh basis carry information on the system dynamics, whereas the non-vanishing diagonal component has information on the state of the field. 

The correlation functions in momentum space  are defined as the Fourier transform of the spacetime correlators with respect to the difference variable $\Delta=x-x'$ keeping constant the central point $X=(x+x')/2$:
\begin{equation} \label{MidPoint}
    G_{ab}(p;X) = \int \ud[4]{x} \expp{-i p \cdot \Delta}
    G_{ab}(X+\Delta/2,X-\Delta/2).
\end{equation}
For homogeneous and static backgrounds the Fourier-transformed propagator does not depend on the central point $X$. The retarded and advanced propagators, which are purely imaginary in the spacetime representation, develop a real part in the momentum representation.

Obviously not all propagators are independent: the complete set of propagators is determined by a symmetric and an antisymmetric function. From the Feynman propagator the other Green functions can be derived, but, in contrast, the retarded propagator lacks the information about the symmetric part of the correlation function. 

\index{Self-energy!in flat backgrounds|(}
\index{Schwinger-Dyson equation}
As in the vacuum, self-energies can be introduced for interacting systems.
The self-energy has a matrix structure and is implicitly defined through the \emph{Schwinger-Dyson equation}:
\begin{equation} \label{SelfEnergyGeneral}
    G_{ab}(x,x') = G_{ab}^{(0)}(x,x')+
    \int \ud[4]{y} \ud[4]{y'} G_{ac}^{(0)}(x,y) [-i\Sigma^{cd}(y,y')]  G_{db}(y',z),
\end{equation}
where $G^{(0)}_{ab}(x,x')$ are the propagators of the free theory, and $G_{ab}(x,x')$ are the propagators of the full interacting theory. The CTP indices $a,b,c\ldots$ are either 1 or 2, and we use a Einstein summation convention for repeated CTP indices. The $ab$ component of the self-energy can be computed in the direct basis, similarly
to the vacuum case, as the sum of all one-particle irreducible
diagrams with  amputated external legs that begin and end with
type $a$ and type $b$ vertices, respectively (cf. appendix \ref{app:CTP}). 

A particularly useful combination is the  retarded self-energy,
defined as $\Sigma_\mathrm R(x,x') = \Sigma^{11}(x,x') +
\Sigma^{12}(x,x')$. It  is related to the retarded propagator through
\begin{equation} \label{SelfEnergyGeneralRetarded}
    \GR(x,x') = \GR^{(0)}(x,x')+
    \int \ud[4]{y} \ud[4]{y'} \GR^{(0)}(x,y) [-i\SigmaR(y,y')]  \GR(y',z),
\end{equation}
This equation can be regarded as a consequence of the causality properties of the retarded propagator. A similar relation holds between the
advanced propagator $G_\mathrm A(x,x')$ and the
advanced self-energy $\Sigma_\mathrm A(x,x') = \Sigma^{11}(x,x') + \Sigma^{21}(x,x')$. Another useful combination is the Hadamard self-energy, which is defined as $\Sigma^{(1)}(x,x') = \Sigma^{11}(x,x') + \Sigma^{22}(x,x')$ [or equivalently as $\Sigma^{(1)}(x,x') =- \Sigma^{12}(x,x') - \Sigma^{21}(x,x')$]   and which is
 related to the Hadamard propagator through \cite{ArteagaThesis}
\begin{equation} \label{GSigmaN}
	G^{(1)}(x,x') = -i \int \ud[4]{y} \ud[4]{y'} \GR(x,y) \SigmaN(y,y') \GA(y',x')
\end{equation}
if the system has a sufficiently dissipative dynamics and interaction is assumed to be switched on in the remote past (otherwise the right hand side of the above equation would incorporate an extra term).
All self-energy combinations can be determined from the knowledge of the Hadamard self-energy and the imaginary part of the retarded self-energy.


So far, all expressions to arbitrary background states $\hat \rho$. For static, homogeneous and isotropic backgrounds,  \Eqref{SelfEnergyGeneralRetarded} can be solved for the retarded propagator by going to the momentum representation:
\begin{equation}\label{GSigmaDiagonal}
	\GR(\omega,\vect p) = \frac{-i}{ -\omega^2 + E_\vect p^2+\SigmaR(\omega,\vect p)-p^0 i\epsilon}.
\end{equation}
Notice that in general the self-energy is a separate function of the energy $\omega$ and the 3-momentum $\vect p$, and not only a function of the scalar $p^2$, as in the vacuum. 
The Hadamard function admits the following expression [which can be derived from \Eqref{GSigmaN}]:
\begin{equation}\label{SelfEnergyHadamardSimp}
	G^{(1)}(\omega,\vect p) = i |\GR(\omega,\vect p)|^2 \SigmaN(\omega,\vect p)  = \frac{i\SigmaN(\omega,\vect p)}
	{ [-\omega^2 + E_\vect p^2+\Re\SigmaR(\omega,\vect p)]^2 + [\Im \SigmaR(\omega,\vect p)]^2}.
\end{equation}
From the retarded propagator and the Hadamard function one can show:
\begin{subequations}\label{PropsSigma}
\begin{align}
    G_\mathrm F(\omega,\vect p)  &= \frac{ -i \left[ -\omega^2 + E_\vect p^2 + \Re \SigmaR(\omega,\vect p)
    \right]
    + i\SigmaN(\omega,\vect p)/2}{\left[ -\omega^2 + E_\vect p^2+ \Re \SigmaR(\omega,\vect p)\right]^2 +
    \left[\Im \SigmaR(\omega,\vect p)\right]^2}, \\
    G_\mathrm D(\omega,\vect p)  &= \frac{ i \left[ -\omega^2 + E_\vect p^2 + \Re \SigmaR(\omega,\vect p)
    \right]
    + i\SigmaN(\omega,\vect p)/2}{\left[ -\omega^2 + E_\vect p^2+ \Re \SigmaR(\omega,\vect p)\right]^2 +
    \left[\Im \SigmaR(\omega,\vect p)\right]^2}, \\
    G_-(\omega,\vect p) &= \frac{i\SigmaN(\omega,\vect p)/2 + \Im \SigmaR(\omega,\vect p)}{\left[ -\omega^2 + E_\vect p^2+ \Re \SigmaR(\omega,\vect p)\right]^2
    +
    \left[\Im \SigmaR(\omega,\vect p)\right]^2}, \\
    G_+(\omega,\vect p) &= \frac{i \SigmaN(\omega,\vect p)/2 - \Im \SigmaR(\omega,\vect p)}{\left[ -\omega^2 + E_\vect p+ \Re \SigmaR(\omega,\vect p)\right]^2
    +
    \left[\Im \SigmaR(\omega,\vect p)\right]^2}.
\end{align}
\end{subequations}
The imaginary part of the self-energy can be interpreted in terms of the net decay rate  $\Gamma_\vect p(\omega)$ for an excitation of energy $\omega$---\ie, decay rate $\Gamma^-_\vect p(\omega)$ minus creation rate $\Gamma^+_\vect p(\omega)$ \cite{Weldon83,Arteaga07b}:
\begin{subequations}
\begin{equation} \label{Weldon}
\begin{split}
	\Im\SigmaR(\omega;\vect p) = -\omega[\Gamma^-_\vect p(\omega) - \Gamma^+_\vect p(\omega)] = -\omega \Gamma_\vect p(\omega)
\end{split}
\end{equation}
In turn the Hadamard self-energy can be interpreted in similar terms as \cite{Arteaga07b}:
\begin{equation} \label{SigmaNCorr}
	\SigmaN(\omega)=-2i|\omega|[\Gamma^-_\vect p(\omega) + \Gamma^+_\vect p(\omega)]= -2i |\omega| \Gamma_\vect p(\omega)[1+2n(\omega)],
\end{equation}
\end{subequations}
where $n(\omega)$ is the occupation number of the modes with energy $\omega$. When the field state is not exactly homogeneous, the expressions in this paragraph are still correct up to order $Lp$, where $L$ is the relevant inhomogeneity time or length scale.

In most sections of this paper we discuss general properties of the propagators and self-energies, which do not depend on any perturbative expansion. In practice, however, many times the only way to evaluate the interacting propagators is through a perturbative expansion in the coupling constant.
For physically reasonable states far ultraviolet modes of the field are not occupied. Therefore, there is an energy scale beyond which the field can be treated as if it were in the vacuum. In the case of thermal field theory this scale is given by the temperature, and  the Bose-Einstein function can be viewed as a natural soft cutoff to the thermal contributions. The counterterms which renormalize the vacuum theory also make the theory ultraviolet-finite in any physically reasonable state. 
This does not mean however that the renormalization process is not modified. As we shall see in the following sections, and as it is shown in Refs.~\cite{DonoghueHolstein83,DonoghueEtAl85,Arteaga07b}, finite parts of the counterterms need be adjusted in a different way for each background if  a physical meaning is to be attributed to the different terms appearing in the action.

The infrared behavior is more subtle. On the one hand, naive perturbation theory may break down under certain circumstances. In the case of thermal field theory, if the relevant contribution to a given process comes from external soft momenta, in order to do perturbation theory in a meaningful way the tree level propagators must be replaced by Braaten and Pisarski's resummed propagator \cite{BraatenPisarski90a,BraatenPisarski90b,Pisarski89}, which incorporates the effect of the hard thermal loops \cite{LeBellac,KraemmerRebhan04}.
On the other hand, even if resummed propagators are used, infrared divergences may still arise when the masses of the fields are negligible. These divergences may show up in the final results of the calculations, or can be hidden in the intermediate stages, leading to finite but incorrect results if they are not properly regularized \cite{Rebhan92}. Additionally, as we shall comment later on, the infrared behavior may lead to a modifications to the ordinary quasiparticle decay law \cite{Weldon02b, BlaizotIancu96,BlaizotIancu97,BoyanovskyEtAl98}. The investigation of the infrared divergences at finite temperature is still an open problem \cite{KraemmerRebhan04}.

\index{Propagator!in flat backgrounds|)}

\section{Open system  viewpoint for the quantum field modes  }

The mode corresponding to the propagating quasiparticle can naturally be regarded as an open quantum system \cite{Davies,BreuerPetruccione,GardinerZoller}: naively, the field mode would constitute the reduced system and all the other modes would form the environment.
However, since the mode-decomposed field operator is a complex quantity obeying the contraint $\phi_\vect p = \phi^*_{-\vect p}$, instead of focusing on a single mode, it proves more useful to choose as the system of interest any two modes with given opposite momentum, and as the environment the remaining modes of the field, as well as the modes of any other field in interaction. 

The field $\phi$ can be decomposed in modes according to
\begin{equation}\label{ModeDecomp}
 	\phi_\vect p = \frac{1}{\sqrt{V}} \int \ud[3]{\vect x}  \expp{-i \vect p \cdot \vect x} \phi(\vect x),
 \end{equation}
where $V$ is the volume of the space, a formally infinite quantity which plays no role at the end. 
Given a particular momentum $\vect p$, the system is composed by the two modes $\phi_\vect p$ and   $\phi_{-\vect p}$, and the environment is composed by the other modes of the field, $\phi_\vect q$, with $\vect q \neq \pm \vect p$. Should there be other fields in interaction of any arbitrary spin, the modes of these additional fields would also form part of the environment.  The entire system is in a state $\hat \rho$; the state of the reduced system is $\hat\rho_\text{s} = \Tr_\text{env} {\hat\rho}$, and the state of the environment is $\hat\rho_\text{e} = \Tr_\text{sys} {\hat\rho}$.  Generally speaking, the state for the entire system is not a factorized product state (\ie, $\hat \rho \neq \hat\rho_\text{s} \otimes \hat\rho_\text{e}$). 
 
 The action can be decomposed as $S = S_\text{sys} + S_\text{count} + S_\text{env} + S_\text{int}$, where $S_\text{sys}$ is the renormalized system action,
\begin{subequations}\label{OQSAction}
\begin{equation}
	S_\text{sys} = \int \ud t  \left( \dot \phi_\vect p \dot \phi_{-\vect p} -E^2_\vect p \phi_\vect p \phi_{-\vect p} \right), \label{SysAction}\\
\end{equation}
 $S_\text{count}$ is the appropriate counterterm action,
\begin{equation} \label{OQSActionEnv}
	S_\text{count}=  
	\int \ud t  \left\{ ({\mathcal Z}_\vect p-1)\dot \phi_\vect p \dot \phi_{-\vect p} -\big[{\mathcal Z}_\vect p (\vect p^2+m_{0}^2)- E^2_\vect p\big] \phi_\vect p \phi_{-\vect p} \right\},
\end{equation}
\end{subequations}
and $S_\text{env}$ and $S_\text{int}$ are respectively the environment and interaction actions.
Notice that we have allowed for an arbitrary rescaling of the field $\phi \to \phi/{\mathcal Z}^{1/2}_\vect p$ and for an arbitrary frequency of the two-mode system $E_\vect p$, which needs not be necessarily of the form $(\vect p^2 + m^2)^{1/2}$.  Since it is always possible to freely move finite terms from the system to the counterterm action and \emph{vice versa}, both the field rescaling and the frequency renormalization should be taken into account even if no infinities appeared in the perturbative calculations. In this paper we will investigate a physical criterion in order to fix the values of these two parameters. Notice that this is not just a matter of notation: in particular, the election of $E_\vect p$ will determine the value of the energy of the quasiparticle. 

The environment and interaction actions  depend on the particular field theory model, and in general not many things can be said about them.
However, provided that the state of the field is stationary, homogeneous and isotropic, under a Gaussian approximation the real environment  can be always equivalently replaced by a one-dimensional massless field and the real interaction  can be replaced by an effective linear interaction with this environment \cite{Arteaga07b}.  In other words, any  scalar two-mode system can be equivalently represented in terms of a pair of quantum Brownian particles 
\cite{GardinerZoller,CaldeiraLeggett83b,UnruhZurek89}, this is to say, by a pair of quantum oscillator interacting linearly with a one-dimensional massless field. 
In this paper we will apply the Gaussian approximation, and it will prove useful for us to reason in terms of the effective coupling constant and the effective environment. The explicit details of the correspondence, which can be found in Ref.~\cite{Arteaga07b}, will not be needed though.

\chapter{Particles in the vacuum} \label{II} 

\index{Propagator!in the vacuum|(}

We begin by reviewing some aspects of the notion of particle in standard quantum field theory in the Minkowski vacuum. While most results in this section can be found in standard quantum field theory textbooks (see for instance Refs.~\cite{BrownQFT,WeinbergQFT,Peskin,ItzyksonZuber,GreinerQFT,Hatfield}), we present them in some detail because, first, analogous steps will be followed when studying quasiparticle excitations in general backgrounds, and, second, in order to clarify some aspects which will prove relevant later on. Except where more detailed references are given, we address the reader to the aforementioned textbooks for the remaining of this section.

 
Let us consider a scalar field theory whose degree of freedom is the scalar field operator $\hat\phi$. 
If the theory is free, the field operators are connected to the creation and annihilation operators via:
  \begin{equation}
 	\hat\phi_\vect p = \frac{1}{\sqrt{2E_\vect p}} \big(\hat a_\vect p + \hat a^\dag_{-\vect p}\big),
 \end{equation}
where $E_\vect p^2= m^2+\vect p^2$ and where  $\hat a^\dag_\vect p$ and $\hat a_\vect p$ are the creation and annihilation operators respectively, which verify the following conmutation relations:
\begin{equation} \label{ConmutRelat}
	[\hat{a}_\vect p,\hat{a}_\vect q] =  [\hat{a}^\dag_\vect p,\hat{a}^\dag_\vect q] = 0, \qquad
	[\hat{a}_\vect p,\hat{a}^\dag_\vect q] =  \delta_{\vect q\vect p}= \frac{1}{V} {(2\pi)^3 \delta^{(3)}{(\vect p - \vect q)}},
\end{equation}
where we recall that $V$ is the (formally infinite) space volume, 
If the theory is interacting, the same above relations hold in the interaction picture. The normalization is chosen so that it closely resembles the quantum mechanical normalization with a finite number of degrees of freedom.

\index{Fock space}

The Hilbert space of the states of the theory has the structure of a Fock space. For non-interacing theories, the Fock space can be built with the aid of the creation and annihilation operators:
	$|\vect p_1\cdots\vect p_n\rangle = \hat a^\dag_{\vect p_1} \cdots \hat a^\dag_{\vect p_n} |0\rangle$
(this  equation assumes that all momenta are different; if this is not the case, the right hand side should incorporate a factor $1/m!$ for each repeated momentum).  The states this way created  have well-defined momentum and energy:
\begin{subequations}
\begin{align}
	\hat{\vect p} |\vect p_1\cdots\vect p_n\rangle &=(\vect p_1 + \cdots + \vect p_n) |\vect p_1\cdots\vect p_n\rangle, \\
	\hat{H} |\vect p_1\cdots\vect p_n\rangle &=(E^{(0)}+E_{\vect p_1} + \cdots + E_{\vect p_n}) |\vect p_1\cdots\vect p_n\rangle,
\end{align}
\end{subequations}
where the momentum and energy operators are:
\begin{align}
	\hat{\vect p} &= \int \udpi[3]{\vect k} \vect k\, \hat a^\dag_\vect k \hat a_\vect k, \qquad
	\hat{H} = \int \udpi[3]{\vect k} E_\vect k \left(  \hat a^\dag_\vect k \hat a_\vect k +\frac12 \right).
\end{align}
 Any state of the field can be expanded in the Fock space: \[|\Psi\rangle = \sum_m \prod_{\vect p_1, \ldots, \vect p_m} f(\vect p_1, \ldots, \vect p_m) |\vect p_1 \cdots \vect p_m \rangle\].
Complementary to the Fock space expansion, the Hilbert space of the field also admits a mode decomposition. Namely, every pure state of the field can be decomposed in the following way:
$
	|\Psi\rangle =  \sum_{n_\vect k} \prod_\vect k c(n_\vect k) |n_\vect k\rangle.
$
where  the modes $|n_\vect k\rangle$ verify
\begin{equation}
	\hat{\vect p} |n_\vect k\rangle = n_\vect k \vect k |n_\vect k\rangle, \qquad \hat{H} |n_\vect k\rangle = n_\vect k E_\vect k |n_\vect k\rangle
\end{equation}

\index{In state@\emph{In} state}
\index{Out state@\emph{Out} state}
\index{Mass!physical}
The situation gets more involved when the theory becomes interacting. In this case, the Hilbert space is still spanned by a set of eigenvectors of the momentum operator $\hat{\vect p}$ and the the full Hamiltonian $\hat H$. However, multiparticle states cannot be labeled by the momentum of each particle in the state, because particles are interacting and the momentum of the particles (and even the number of them) changes. However, in the remote future and past particles are well separated and the interaction between them is negligible.
 Labeling by $|\vect p_1\cdots\vect p_n\rangle_\text{in(out)}$  the eigenstate of the full Hamiltonian that corresponds to a multiparticle state in the limit $t\to-\infty$ ($t\to\infty$), one has
\begin{equation}
	\hat{H} |\vect p_1\cdots\vect p_n\rangle_\text{in(out)} =(E_{\vect p_1} + \cdots + E_{\vect p_n}) |\vect p_1\cdots\vect p_n\rangle_\text{in(out)},
\end{equation}
where in this case $E_\vect p^2 = \mph^2 + \vect p^2$, with $\mph$ being the \emph{physical mass} of the particles, which differs in general from the bare mass present in the Lagrangian. Notice that the particles appearing in the \emph{in} or \emph{out} states do not necessarily correspond to the particles appearing in the corresponding free theory: any unstable particle will not appear in the asymptotic states, and we will possibly have to add bound states to the asymptotic states. (For simplicity our notation assumes just one particle species and does not take into account these possibilities.) Notice also that the \emph{in} and \emph{out} states are defined for all times (although they only have special properties in the asymptotic limits). Therefore, either the \emph{in} or \emph{out} Fock spaces built from those states can be chosen as a basis for the Hilbert space of the interacting theory. In the remote past and future one can build a free theory that matches the properties of the interacting theory in these regimes. These ``free states'' correspond unitarily to the \emph{in} and \emph{out} states of the interacting theory in the asymptotic limits\footnote{In order to properly define this correspondence one must work with wavepackets, so that there is localization in time and space; otherwise the states of the Fock space are completely delocalized. See Ref.~\cite{WeinbergQFT} for more details.} (the correspondence being different in each case). The in and out states are therefore also unitarily related (via the S matrix). See Refs.~\cite{Haag,GreinerQFT,WeinbergQFT} for more details.

\section{Particles: the spectral approach}
\subsection{The K\"all\'en-Lehmann spectral representation}

\index{Propagator!Whightman}
\index{Spectral function}
\index{K\"all\'en-Lehmann spectral representation|see{spectral representation}}
\index{Spectral representation}

An arbitrary correlation function $G_\text{any}(x,x')$ admits the following K\"all\'en-Lehmann spectral representation,
\begin{align}
	G_\text{any}(x,x') &= \int_0^\infty \ud s \rho(s) G_\text{any}\fr(x,x';s),
\end{align}
where   $G_\text{any}\fr(x,x';s)$ is the corresponding free  function with squared mass parameter $s$ and  $\rho(s)$ is the \emph{vacuum spectral function}, which is defined as
\index{Spectral function} 
\begin{equation}\label{SpectralFunction}
	\rho(-p^2) := 
\frac{1}{2\pi} \sum_\alpha (2\pi)^4 \delta^{(4)}(p-p_\alpha) |\langle 0 |\phi(0)|\alpha\rangle|^2,
\end{equation}
and which verifies, among others, the following properties:
(i) $\rho(s) \geq 0$, (ii) $\rho(s) = 0$ for $s<0$ and (iii) $\int_0^\infty \ud{s} \rho(s) = 1$.
In this last equation  $|\alpha\rangle$  is a complete set of orthonormal states, eigenstates of the four-momentum operator $\hat p = (\hat H,\hat {\vect p})$, spanning the identity:
$
	\hat 1 = \sum_\alpha |\alpha\rangle\langle \alpha|$,	$\hat p^\mu |\alpha\rangle = p^\mu_\alpha |\alpha\rangle$.
An example of such class of states are the \emph{in} or \emph{out} states just considered.

The spectral representation can be also expressed in  momentum space:
\begin{subequations}
\begin{align}
	G_+(p) &= 2\pi \rho(-p^2) \theta(p^0), \\
	G(p) &= 2\pi \rho(-p^2) \sign(p^0), \label{PauliJordanSpectral}\\
	\GR(p) &= \int_0^\infty \frac{-i\rho(s)\,\vd s}{-(p^0+i\epsilon)^2+\vect p^2+s},\label{retardedSpectral} \\
	\GF(p) &= \int_0^\infty \frac{-i\rho(s)\,\vd s}{p^2+s-i\epsilon}. \label{FeynmanSpectral}
\end{align}
\end{subequations}
The first  two equations show that in vacuum the Whightman functions and the Pauli-Jordan propagator essentially amount to the spectral function. The latter two equations show that the retarded propagator and the Feynman propagator have well-defined analyticity properties when considered functions in the complex $p^2$ plane. In fact, they also have analyticity properties in the complex energy plane, as shown in the following two equivalent representations:
\begin{subequations}
\begin{align}
	\GR(p) &= \int \ud {k^0} \left[ \frac{i \rho(k^{02} - \vect p^2 )}{p^0 - {k^0} + i\epsilon} -  \frac{i \rho(k^{02} - \vect p^2 )}{p^0 + {k^0} + i\epsilon} \right], \\
	\GF(p) &= \int \ud {k^0} \left[ \frac{i \rho(k^{02} - \vect p^2 )}{p^0 - {k^0} + i\epsilon} -  \frac{i \rho(k^{02} - \vect p^2 )}{p^0 + {k^0} - i\epsilon} \right] 
\end{align}
\end{subequations}
Taking into account the relation of the spectral function with the Pauli-Jordan propagator, given by Eq.~\eqref{PauliJordanSpectral}, one can also write
\begin{equation}\label{SpectralVacuumAlt}
	\GR(p) = \int \frac{\vd k^0}{2\pi}   \frac{i G(k^0,\vect p)}{p^0 - k^0 + i\epsilon}.
\end{equation}
This last equation also follows directly from the definition of the retarded propagator. 


\subsection{Stable particles}

From all the states of the theory, let us single out the one-particle states corresponding to stable particles (assuming they exist), characterized by the momentum $\vect p$ and the physical mass $\mph$: $\hat{\vect p }|\vect p\rangle = \vect p |\vect p\rangle$, $\hat{H }|\vect p\rangle = E_\vect p|\vect p\rangle$, with $E_\vect p^2 = \vect p^2 + \mph^2$.  Considering those states explicitly, the spectral function can be developed as
\begin{equation}
	\rho(-p^2) = Z \delta(p^2 + \mph^2) + \theta(p^2+m_*^2) \sigma(-p^2) ,
\end{equation}
where $	Z=\abs{\langle 0 |\phi(0)|\vect p\rangle}^2 $
is a positive constant (which does not depend on the momentum because of the Lorentz invariance of the theory), $\sigma(-p^2)$ is the contribution from the multiparticle states, and $m_*$ is the minimum rest energy of the multiparticle states. The $Z$ constant is frequently renormalized to one by rescaling the field (which amounts to adding a suitable counterterm to the original action). 

\index{Propagator!analytic structure}

According to Eq.~\eqref{FeynmanSpectral}, the Feynman propagator in momentum space has the structure:
\begin{equation}
	\GF(p) = \frac{ -i Z}{p^2 + m^2 - i\epsilon} + \int_{m_*^2}^\infty \vd s \frac{ -i \sigma(s)}{p^2 + s - i\epsilon}.
\end{equation}
Thus, the Feynman propagator is analytic in the complex $p^2$ plane, except for a pole at $p^2 = -\mph^2 + i\epsilon$, and a branch cut starting at $p^2 = -m_*^2 + i\epsilon$ and running parallel to the real axis. There are additional poles if bound states can be formed; we shall not take into account this possibility in the following. In presence of massless particles the pole and the cut may overlap. When doing perturbation theory this is reflected in the appearance of infrared divergences in the calculation of the residues of the pole. For a discussion of this point in the case of QED see \eg\ Ref.~\cite{Hatfield}. We will assume in the following that the pole and the cut are well-separated.


\index{Propagator!analytic structure}

\index{Self-energy!in the vacuum}
In perturbative field theory, the renormalized Feynman propagator can be resummed in the following way
\begin{equation}\label{SelfEnergyVacuum}
	\GF(p) = \frac{-i}{p^2 + m^2 + \Sigma(-p^2)},
\end{equation}
where $m$ is the renormalized mass and $\Sigma(-p^2)$ is the \emph{self-energy}, corresponding to the sum of all one-particle irreducible Feynman diagrams. The locus of the poles and cuts is given by the solution of the equation
$
	p^2 + m^2 + \Sigma(-p^2) = 0. 
$
The zeros of this equation lie next to the real axis as dictated by the spectral representation \eqref{FeynmanSpectral}. The lowest zero of the equation is the pole at $p^2=-\mph^2$. 
\index{On-shell renormalization condition}
With the on-shell renormalization conditions the renormalized mass coincides with the physical mass, $m=\mph^2$, so that $\Sigma(-\mph^2)=0$.

\index{Propagator!analytic structure}

\begin{figure}
\centering
\includegraphics[width=.7\textwidth]{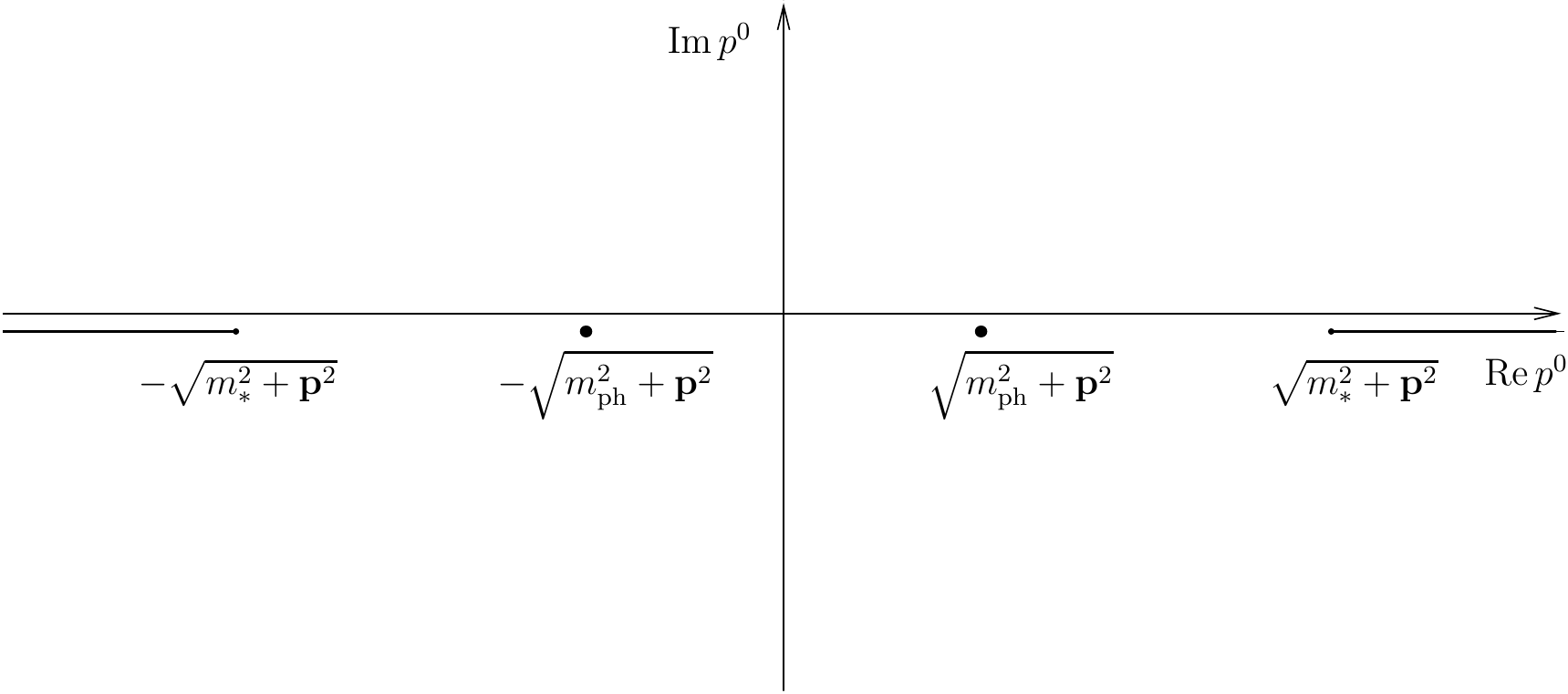}
\caption{Analytic structure of the retarded propagator in the vacuum, as seen in the complex energy plane. There are two poles corresponding to the stable particle, and two branch cuts, whose branching points indicate the minimum energy for the multiparticle states. Between the poles and the branching points there might be as well other poles corresponding to bound states (not shown in the plot).}
\label{fig:retardedVacuum}
\end{figure}

Although it is usual to consider the Feynman propagator, the analytic structure of other propagators can be studied as well. The retarded propagator will be of special importance:
\begin{equation} \label{VacuumRetardedAnalytic}
	\GR(p) = \frac{ -i Z}{p^2+ m^2 - i\epsilon p^0} + \int_{m_*^2}^\infty \vd s \frac{ -i \sigma(s)}{p^2 + s - i\epsilon p^0}.
\end{equation}
As shown in figure \ref{fig:retardedVacuum}, in the complex energy plane (complex $p^0$ plane), the retarded propagator is analytic except for the two poles at $p^0 = \pm E_\vect p - i \epsilon $ and the two branch cuts.

\subsection{Unstable particles}\label{sect:Unstable}

\index{Particle!unstable}

Recall that unstable particles are not asymptotic states and hence do not correspond to eigenstates of the Hamiltonian nor they belong to the 1-particle sector of the Fock space. Rather, they are combinations of those multiparticle states corresponding to the decay products of the unstable particle. In terms of the K\"all\'en-Lehmann representation of the propagator, this means that unstable states are represented by a branch cut rather than a pole. Hence the spectral function for the multiparticle states is given by 
$
	\rho(-p^2) = \theta(p^2+m_*^2) \rho(-p^2) ,
$
 where $m_*$ is in this case the minimum energy to create a multiparticle state.

Unstable particles however develop a pole of the propagator in a second Riemann sheet of the complex $p^2$ plane \cite{MatthewsSalam58a,MatthewsSalam58b,BrownQFT,JacobSachs60,Veltman63,Cocolicchio98,GalindoPascual}. The Feynman propagator can be resummed as
\begin{equation}
	\GF(p) = \frac{-i}{p^2+m^2 + \Sigma(-p^2) } = \frac{-i}{p^2+m^2 + \Re \Sigma(-p^2) + i\Im \Sigma(-p^2) }.
\end{equation}
We now define the mass of the unstable particle as the lowest order solution of
\begin{equation}
	-\mph^2+ m^2 + \Re \Sigma(\mph^2) = 0
\end{equation}
and identify
\begin{equation}
	\gamma =-\frac{1}{\mph} \Im \Sigma(\mph^2)
\end{equation}
as the decay rate in the particle rest frame, according to the optical theorem. The mass as defined above corresponds approximately to the rest energy of the particle (this assertion will be checked later on), although it should be noted that the energy of an unstable particle fluctuates according to the time-energy uncertainty principle. 
Thus, the second Riemann sheet of the Feynman propagator in momentum space has a pole in the region $\Im \Sigma(p^2) < 0$, whose real part corresponds to the approximate mass of the particle and whose imaginary part corresponds to the decay rate:
\begin{equation}\label{FeynmanUnstable}
	\GF(p) \approx \frac{-iZ}{p^2+\mph^2 - i \mph \gamma} + \widetilde G(p),
\end{equation}
The function $ \widetilde  G(p)$ is analytic function in the vicinity of the pole, but it does develop a singular behavior when approaching the different particle creation thresholds.

\section{Particles: the real-time approach}\label{sect:TimeEvol}

\subsection{Time-evolution of the propagators}

So far we have carried out the analysis in the energy-momentum representation. It will prove also illustrative to consider the time-momentum representation of the propagator,
\begin{equation*}
	\GF(t,t';\vect p) = \int\frac{\vd p^0}{2\pi} \expp{-ip^0(t-t')} \GF(p).
\end{equation*}
The aim is to compute the behavior of the propagator for large time lapses. We shall consider both the stable and unstable cases simultaneously (with $\gamma=\epsilon$ if the particle is stable). 

From \Eqref{FeynmanUnstable}, the time behavior of the pole can be easily derived:
\begin{equation*}
	\GF(t,t';\vect p) = \frac{-iZ}{2\sqrt{\vect p^2+ \mph^2-i\mph\gamma}} \expp{-i\sqrt{\vect p^2 + \mph^2 - i \mph \gamma} |t-t'|} + \widetilde G(t,t';\vect p)
\end{equation*}
In order for the particle concept to be meaningful, the condition $\gamma \ll \mph$ must be verified (otherwise one would speak of wide resonances rather than particles). Under these conditions, the above expression can be approximated as
\begin{equation}\label{GFTimeDecay}
	\GF(t,t';\vect p) = \frac{-iZ}{2E_\vect p} \expp{-iE_\vect p|t-t'|}\expp{- \Gamma_\vect p |t-t'|/2} + \widetilde G(t,t';\vect p)
\end{equation}
with
$
	E^2_\vect p = \vect p^2 + \mph^2,
$
and where we have defined the decay rate in the laboratory rest frame
\begin{equation}
	\Gamma_\vect p := \frac{ m \gamma}{E_\vect p} = -\frac{1}{E_\vect p} \Im \Sigma(\mph^2).
\end{equation}
\index{Decay rate}

The behavior of remaining piece $\widetilde G(t,t';\vect p)$ remains to be determined. In general, it depends on the precise value of the spectral function along the branch cut, which in turn depends on the multiparticle structure of the theory. However, by appealing to the Riemann-Lebesgue theorem, one can show \cite{BrownQFT} that an order of magnitude estimation of $\widetilde G(t,t';\vect p)$ is given by
\begin{equation}
	|\widetilde G(t,t';\vect p)|  \sim \frac{1}{M^{\alpha+1} |t-t'|^\alpha},
\end{equation}
where $M$ is the scale in which the leading multiparticle threshold starts, and where $\alpha$ is some positive coefficient which depends on the particular structure of the theory.

For unstable particles three different time regimes can be therefore distinguished.  For short times, of the order of the de Broglie size of the particle $\mph^{-1}$, transient effects dominate, and the  behavior depends on the particular details of the field theory model.  For large time lapses ($|t-t'|\gg M^{-1}$), the  behavior is dominated by the pole, which in this case is located off the real axis. The modulus of the propagator decreases exponentially with a rate $\Gamma_{\vect p}$. Transient effects are subdominant in this regime, since they decay in a much faster timescale $M^{-1}$. For extremely long times, however, ``transient'' effects dominate again, since any power low decay dominates over an exponential decay for sufficiently long times.  Experimentally the breakdown of the exponential law for very long times is almost never observable, since power-law terms dominate again after many particle lifetimes, when the chances of observing the particle are almost null (as we will see in the following).


\subsection{Two-point functions and asymptotic fields}\label{sect:Asympt}

\index{Asymptotic fields}


We have just seen that the 2-point correlation functions match to the 2-point correlation function of a free field plus an additional multiparticle contribution, which vanishes for long times:
\begin{subequations}
\begin{equation}
	G_+(t,t';\vect p) = Z G_+^\text{(1p)}(t,t';\vect p) + \widetilde G_+(t,t';\vect p) \xrightarrow[t-t'\to\infty]{} Z G_+^\text{(1p)}(t,t';\vect p)
\end{equation}
or equivalently
\begin{equation}
	\langle 0 | \hat\phi_\vect{p} U(t,t') \hat\phi_\vect{-p} |0 \rangle  \xrightarrow[t-t'\to\infty]{}  \frac{Z}{2E_\vect p} \, \langle \vect p| U(t,t') | \vect p\rangle
\end{equation}
\end{subequations}
where $|0\rangle$ and $|\vect p\rangle$ are the vacuum and 1-particle state of the interacting theory respectively, and where $U(t,t')$ is the time evolution operator.
 Therefore, we can make the identification
\begin{equation}\label{PField}
	|\vect p\rangle \cong \frac{\sqrt{2 E_\vect p}}{\sqrt{Z}} \hat\phi_{-\vect p} | 0\rangle .
\end{equation}
The symbol $\cong$ here means equivalence when evaluated in a matrix element in the limit of large time lapses.  Physically, the field operator excites the one-particle state and the multiparticle sector, but multiparticle excitations are off the mass shell and they decay quickly.

\index{In state@\emph{In} state}
\index{Out state@\emph{Out} state}
\label{Asymptotic field operator}
The heuristic argument given above can be connected to the fact that the particle content of the theory corresponds to that of a free theory in the asymptotic limits.  Let us consider the asymptotic field operator $\phir$,  which by assumption obeys free equations of motion,
\begin{subequations}
\begin{equation} \label{EffeciveFree}
	(\partial_\mu\partial^\mu + \mph^2) \phir =0, \qquad
	- \ddot \phir_\vect p + (\vect p^2+\mph^2)\phir_\vect p=0,
\end{equation}
and which corresponds to the field operator through
\begin{equation}
	\phir(\vect x) \cong   Z^{-1/2} \phi(\vect x), \qquad
	 \phir_\vect p \cong Z^{-1/2} \phi_\vect p.
\end{equation}
\end{subequations}
Let us also consider the corresponding creation and annihilation operators,
\begin{align} \label{46}
	\hat \ar_\vect p &= \sqrt{2E_\vect p} \left[ \hat\phir_{-\vect p} +   \frac{i}{E_\vect p} \hat\pir_{-\vect p} \right], \qquad
	\hat \ar^\dag_\vect p = \sqrt{2E_\vect p} \left[ \hat \phir_{\vect p} -   \frac{i}{E_\vect p} \hat \pir_{\vect p} \right],
\end{align}
where $\hat \pir_\vect p$ is the canonical momentum operator and $E_\vect p^2=\vect p^2 + \mph^2$. \emph{In} and \emph{out} Fock spaces can be constructed with the field $\phir$ \cite{GreinerQFT,Haag,WeinbergQFT}.\footnote{One possible concern is the fact that there is an apparent contradiction between the two canonical conmutation relations $[\hat\phi,\hat\pi]=i$ and $[\hat\phir,\hat\pir]=i$. In reality there is no such contradiction because the two fields are equivalent in the asymptotic limits only when evaluated in matrix elements (weak operator convergence) \cite{GreinerQFT}.  In any case, the reader must be warned  that a fully rigorous analysis is usually not feasible for standard field theories.}
With the asymptotic field operator the one-particle state can be represented as
\begin{equation}\label{PFieldAsym}
	|\vect p\rangle \cong  {\sqrt{2 E_\vect p}} \hat\phir_{-\vect p} | 0\rangle_\text{in/out} =\hat\ar^\dag_\vect p | 0\rangle_\text{in/out} 
\end{equation}
Comparing \Eqref{PField} and \Eqref{PFieldAsym} we see that  the distinction between the  field $\phi$ and the the asymptotic field $\phir$ may be blurred provided the constant $Z$ is renormalized to one and assumes that in the asymptotic limit the field obeys effective free equations of motion. Under these assumptions one may simply write
\begin{equation}\label{PFieldApprox}
	|\vect p\rangle \cong  {\sqrt{2 E_\vect p}} \hat\phi_{-\vect p} | 0\rangle_\text{in/out} \cong \hat a^\dag_\vect p | 0\rangle_\text{in/out} 
\end{equation}

When particles are unstable the situation is less clear since strictly speaking a one-particle sector which also is an eigenstate of the Hamiltonian does not exist  (the only eigenstates being the multiparticle states corresponding to the decay products of the unstable particles). However, if particles are long-lived one may think of approximate 1-particle states (which in fact correspond to multiparticle state combinations). We  assume that those are also approximately given by \Eqref{PFieldApprox}.

\subsection{Physical interpretation of the time representation} 

\index{Propagator!time representation}

We have previously seen that in frequency space the poles of the propagator are connected with the energy and lifetime of the particle.
Let us confirm this statement studying the time evolution of the expectation value of the Hamiltonian operator.  
In field theory, strictly speaking, only asymptotic properties are completely well-defined. However, as long as  sufficiently large time lapses  are considered (compared to the relevant inverse energy scales), approximate results for finite times can be obtained. We shall use the asymptotic particle representation in terms of fields (described in the previous subsection) for finite but sufficiently long time lapses, shall blurry the distinction between the usual and asymptotic fields (supposing that the quantity $Z$ has been renormalized to one) and shall treat the stable and unstable cases simultaneously.

According to \Eqref{SysAction}, the reduced Hamiltonian operator for the relevant two-pair mode is given by\
\begin{equation}\label{HSys}
	\hat H_\text{sys} = E_\vect p \left( \hat{a}^{\dag}_\vect p \hat{a}_\vect p + \hat{a}^{\dag}_{-\vect p} \hat{a}_{-\vect p}  +1 \right).
\end{equation}
If in the remote past time $t_{0}$ a particle is introduced into the system, the state for later times will be $|\vect p\rangle \cong a^\dag_{\vect p}|0\rangle$. The evolution of the number of particles in this state is given by
\begin{equation}
	E(t,t_{0};\vect p) := \langle \vect p|\hat H_{\vect p}(t)|\vect p\rangle = E_\vect p 
		\langle 0|\hat a_{\vect p}U(t_{0},t)  \left(  \hat a^\dag_{\vect p}\hat  a_{\vect p} + \hat{a}^{\dag}_{-\vect p} \hat{a}_{-\vect p}+1\right) U(t,t_{0}) \hat a^\dag_{\vect p}|0\rangle.
\end{equation}
Introducing a resolution of the identity we obtain:
\begin{equation*}
	E(t,t_{0};\vect p) ={E_\vect p}  \sum_{\alpha}		\langle 0|\hat a_{\vect p}U(t_{0},t) \hat a^\dag_{\vect p} |\alpha \rangle \langle \alpha  |\hat  a_{\vect p}U(t,t_{0}) \hat a^\dag_{\vect p}|0\rangle + E^{(0)},
\end{equation*}
where $E^{(0)} =\langle 0 | \hat H_\text{sys} |0 \rangle$ is the vacuum energy. By energy and momentum conservation, only the vacuum survives from the above summation. Therefore:
\begin{equation*}
\begin{split}
	 E(t,t_{0};\vect p) &= E^{(0)}+ {4E_\vect p^3} |\langle 0|\hat \phi_{-\vect p} U(t_{0},t) \hat\phi_{\vect p} |0 \rangle|^2,
\end{split}
\end{equation*}
from where we obtain
\begin{equation}
	 E(t,t_{0};\vect p) =  E^{(0)} + {4E_\vect p^3} |\GF(t_0,t;\vect p)|^2.
\end{equation}
Introducing the explicit value of the propagator, given by \Eqref{GFTimeDecay}, and neglecting the off-shell contribution $\widetilde G(t,t';\vect p)$ we get the expected result
\begin{equation}
	E(t,t_0;\vect p) = E^{(0)}+ E_\vect p \expp{-\Gamma_\vect p(t-t_0)} .
\end{equation}
Particles have energy $E_\vect p$ and decay in  a timescale $\Gamma_\vect p$ in the e domain of validity of the exponential law. 

We recall once more that unstable particles do not correspond to any eigenstate of the Hamiltonian, and thus no asymptotic states can be associated to them. Therefore, strictly speaking, one-particle states are a linear combination of many multiparticle states and one cannot make reference to energy conservation, since energy conservation is associated to asymptotic properties. However, if the lifetime of the particles is long enough one can think of approximate asymptotic states and approximate energy conservation, so that the above calculation would be approximately valid.

\chapter{Quasiparticles: the spectral approach} \label{III}

\section{K\"all\'en-Lehmann spectral representation}

In a general background the quickest and clearest way to derive the spectral representation is by simply recalling the definition of the propagators. From \Eqref{RetardedProp} we immediately obtain the spectral representation for the retarded propagator \cite{ChouEtAl85}
\begin{subequations}
\begin{equation}
	\GR(p;X) = \int \frac{\vd k^0}{2\pi}   \frac{i G(k^0,\vect p;X)}{p^0 - k^0 + i\epsilon}.
\end{equation}
This equation is identical to its vacuum counterpart, \Eqref{SpectralVacuumAlt}. The advanced propagator follows a similar representation,
\begin{equation}
	\GA(p;X) = \int \frac{\vd k^0}{2\pi}   \frac{i G(k^0,\vect p;X)}{p^0 - k^0 - i\epsilon}.
\end{equation}
\end{subequations}
Hereafter the Pauli-Jordan function will also be called \emph{spectral function}. The similarities with the vacuum case end here. The retarded and advanced propagators are the only propagators that can be expressed in terms of an integral of the spectral function.  Notice also that in general the spectral representation  can only be expressed as an integral over the energy, and not as an integral over the invariant mass. 

The Pauli-Jordan function verifies the following properties:
\begin{subequations}
\begin{align}
	  G(p;X) &> 0, \quad \text{if}\ p^0>0,\\
	 G(-p;X) &= -G(-p;X), \\
	\int \frac{\vd k^0}{2\pi} k^{0} G(k;X)&=1.
\end{align}
\end{subequations}
The first two properties are a simple consequence of the definition of the propagator. The third property is a sum rule, consequence of the equal-time commutation relations, 
$
	[\hat\phi(t,\vect x),\partial_t{\hat\phi}(t,\vect y)] = i \delta^{(3)}(\vect x - \vect y).
$

For stationary backgrounds an explicit representation for the Pauli-Jordan function can be obtained, similarly to \Eqref{SpectralFunction}. The Pauli-Jordan function can be expressed in the basis of eigenstates of the Hamiltonian as 
\begin{equation*}
\begin{split}
	G(t,t';\vect p) &= \sum_{\alpha}  \rho_{\alpha} \langle\alpha|[\hat\phi_{-\vect p}(t),\hat\phi_{\vect p}(t')]|\alpha\rangle = \sum_{\alpha}  \rho_{\alpha} \langle\alpha|\hat\phi_{-\vect p} \expp{-i \hat H(t-t')} \hat\phi_{\vect p} \expp{i\hat H(t-t')}|\alpha\rangle - \text{(c.c)},
	 \end{split}
\end{equation*}
where $\rho_{\alpha} = \langle \alpha|\hat\rho|\alpha\rangle$. We have used the fact that the state is stationary, so that the density matrix operators diagonal in the basis of eigenstates of the Hamiltonian. Let us now introduce a resolution of the identity $1=|\beta\rangle\langle\beta|$:
\begin{equation*}
\begin{split}
	G(t,t';\vect p)
	 &= \sum_{\alpha,\beta}  \rho_{\alpha} \langle\alpha|\hat\phi_{-\vect p} \expp{-i \hat H(t-t')} |\beta\rangle \langle\beta|\hat\phi_{\vect p} \expp{i\hat H(t-t')}|\alpha\rangle - \text{(c.c.)}\\
	 &= \sum_{\alpha,\beta}  \rho_{\alpha} \langle\alpha|\hat\phi_{-\vect p}  |\beta\rangle \langle\beta|\hat\phi_{\vect p}|\alpha\rangle \expp{-i(E_{\beta}-E_{\alpha})(t-t')}- \text{(c.c.)}\\
	 &= \sum_{\alpha,\beta} \rho_{\alpha}\abs{\langle\alpha|\hat\phi_{\vect p}|\beta\rangle}^{2} (-2i) \sin{[(E_{\beta}-E_{\alpha})(t-t')]},
	 \end{split}
\end{equation*}
where we have used that $\hat\phi_{\vect p}^\dag = \hat\phi_{-\vect p}$
Going to the frequency space we obtain the desired expression:
\begin{equation} \label{SpectralDetailed}
	G(p) = \sum_{\alpha,\beta} \rho_{\alpha}\abs{\langle\alpha|\hat\phi_{\vect p}|\beta\rangle}^{2} \left[ \delta(p^0 + E_{\alpha} - E_{\beta})-\delta(p^0 - E_{\alpha} + E_{\beta})\right].
\end{equation}
For stationary background states the Pauli-Jordan propagator is proportional to the probability for the field operator of momentum $\vect p$ to induce a transition to a state with higher energy $p^0$, minus the probability to induce a transition to a state of lower energy $p^0$.


\index{Renormalization}

\section{Spectral analysis of the quasiparticles}

\index{Spectral representation}
\index{Spectral function}

Let us assume that around some range of energies the Pauli-Jordan propagator has the following structure:
\begin{equation}\label{StableQuasiparticle} 
	G(\omega,\vect p) = \frac{Z_{\vect p}}{2E_{\vect p}}\big[\delta(\omega-E_\vect p)-\delta(\omega+E_\vect p)], \quad \text{for }|\omega|\sim E_\vect p.
\end{equation}
This means that the field operator with momentum $\vect p$ creates an excitation whose energy is exactly $E_\vect p$. Since there is no spread in the energy, the excitation must be infinitely lived (must be stable). The corresponding retarded propagator is:
\begin{equation} 
	\GR(\omega,\vect p) = \frac{-iZ_{\vect p}}{-\omega^2 +E_\vect p^2 - i\omega\epsilon} , \quad \text{for }|\omega|\sim E_\vect p.
\end{equation}
The excitation this way created would have exact energy $E_\vect p$ and exact momentum $\vect p$. Therefore it corresponds to a stable quasiparticle. Notice that, in contrast to the stable particles in the vacuum, $Z_\vect p$ can depend on the 3-momentum $\vect p$ and $E_\vect p^2$ need not be of the form $\vect p^2+m^2$.

However, stable quasiparticles are an idealization, and do not correspond exactly to any physical situation, since dissipation is a generic feature of non-vacuum states as long as there is interaction. We recall that by ``dissipation'' we do not necessarily mean the quasiparticle decaying into a product of different quasiparticles: ``dissipation'' can simply mean the quasiparticle changing its momentum. In any case, stable quasiparticles may be a good approximation in many situations in which the effective coupling to the environment is very small (in the sense of the quantum Brownian motion correspondence explained in Ref.~\cite{Arteaga07b} and summarized in the introduction). However, this does not mean at all that the real coupling must be small, or that the approximation is limited to weakly interacting systems; as a matter of fact, in strongly coupled situations there might be situations in which assuming free quasiparticles might well be a good approximation. 

\index{Propagator!analytic structure}
Anyway, quasiparticles are in general not stable, and instead decay with some rate $\Gamma_{\vect p}$. The generic form of the retarded propagator in terms of the self-energy is
\begin{equation}
	\GR(\omega,\vect p) = \frac{-i}{-\omega^2 +E_{\vect p}^2 + \SigmaR(\omega,\vect p)}.
\end{equation}
Following the vacuum analysis of subsection \ref{sect:Unstable}, whenever there is a long-lived quasiparticle the analytic continuation of the retarded propagator can be approximated by:
\begin{equation}\label{AnalyticStandard}
	\GR(\omega,\vect p) = \frac{-i Z_\vect p}{-\omega^2 +E^2_{\vect p} - i \omega \Gamma_{\vect p}} + \widetilde G(\omega,\vect p),
\end{equation}
where $\widetilde G(\omega,\vect p)$ is an analytic function in the vicinity of the pole. This means that for unstable (but long-lived) quasiparticles, the Pauli-Jordan function, instead of being a delta function around the energy of the pole, it is an approximate Lorentzian function, whose width corresponds to the decay rate. The momentum-dependent functions $E_\vect p$ and $\Gamma_\vect p$ are given, in terms of the self-energy, by
\begin{subequations} \label{RGammaBasic}
	\begin{align}
		E_\vect p^2 &= m^2 + \vect p^2  + \Re \SigmaR(E_\vect p,\vect p),\\
		\Gamma_\vect p &= - \frac{1}{E_\vect p} \Im \SigmaR(E_\vect p,\vect p),
	\end{align}
\end{subequations}
under the assumption that $\Gamma_\vect p$ is much smaller than $E_\vect p$. The spectral representation indicates that $E_\vect p$ corresponds to the quasiparticle energy. The physical meaning of $\Gamma_\vect p$ can be extracted from the interpretation of the imaginary part of the self-energy, \Eqref{Weldon}: it corresponds to the net decay rate of the quasiparticle.

\index{Propagator!analytic structure}

\begin{figure}
\centering\includegraphics[width=.75\textwidth]{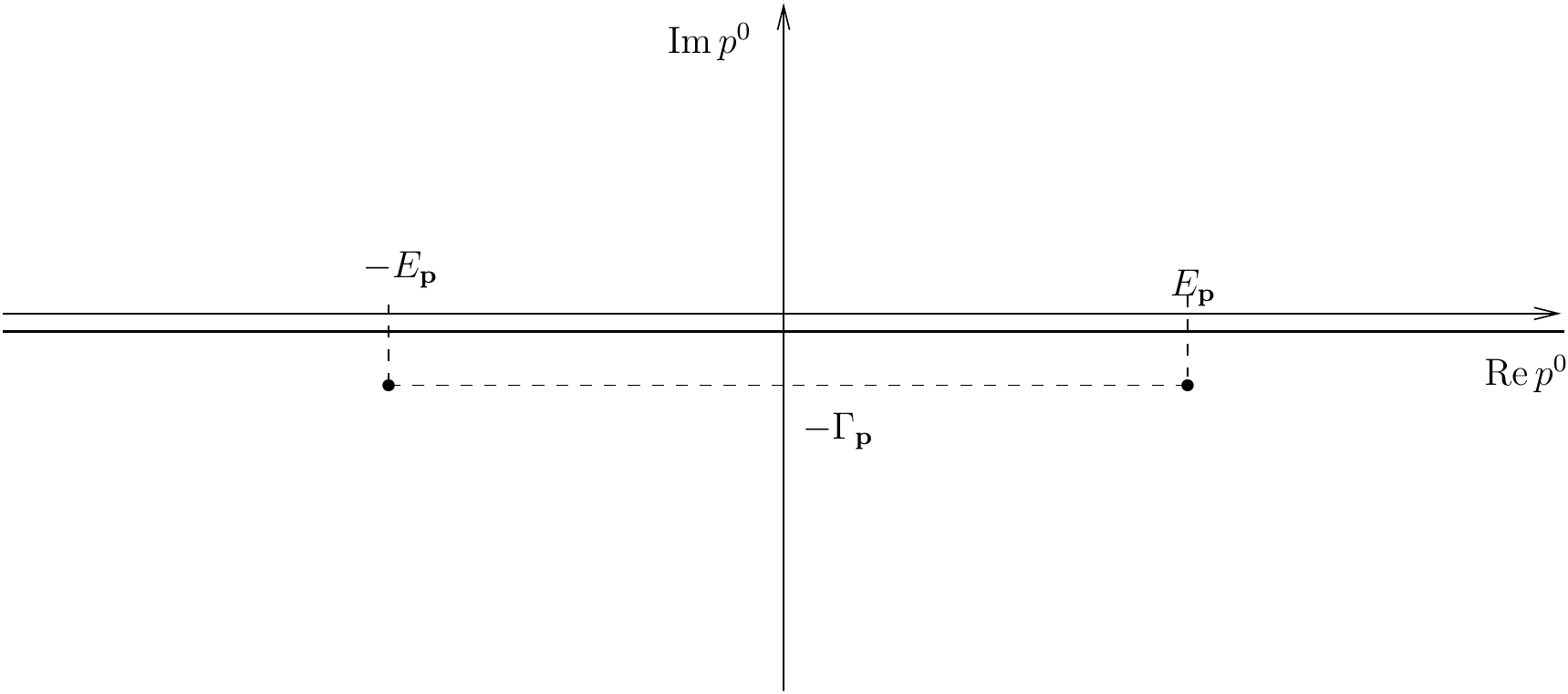}
\caption{Analytic structure of the retarded propagator for quasiparticles in general backgrounds, in the complex energy representation.  The retarded propagator has no poles but only has a cut parallel to the real axis. Its analytic continuation (or second Riemann sheet) has a pair of poles, whose real part corresponds to the energy of the excitation and whose imaginary part corresponds to the decay rate. The cut need not extend from $-\infty$ to $+\infty$: it might be interrupted for some energy sectors.}\label{fig:retarded-alt}
\end{figure}

The structure of the retarded propagator \eqref{AnalyticStandard} is completely analogous to that of the vacuum (see Sec.~\ref{II} and figure \ref{fig:retarded-alt}). There are however two important differences that it is worth commenting. First, in the vacuum the analysis can be equivalently performed with the Feynman or the retarded propagators, while in generic backgrounds the spectral analysis can only be applied to the retarded propagator, since in  general  the Feynman propagator has an additional explicit dependence on the background state of the field. Second, in the vacuum one can study the spectral structure either in terms of the energy or in terms of the squared four-momentum. In general,  there is a preferred reference frame and therefore the explicit Lorentz invariance of the results is broken, and, hence, only a spectral analysis based on the energy is meaningful. See Refs.~\cite{Weldon99,Weldon02} for a more in depth investigation of the analytic structure of the propagators in thermal field theory.

It must be noted that in presence of massless particles, there are situations in which the decay rate appears to be infrared-divergent from a perturbative calculation. This can be attributed to the fact that in presence of massless particles  the retarded propagator does not necessarily exhibit the form \eqref{AnalyticStandard}, because there is no threshold for the creation of massless excitations.  For instance, when computing the lifetime of a quasiparticle in a very hot QED plasma it is found that  the retarded propagator does not exhibit any singularity near the particle resonance energy  \cite{BlaizotIancu96,BlaizotIancu97,Weldon02}, although the propagator is still strongly peaked around this point.  Although we will  comment further on this possibility, in the following we will limit our calculations to those situations in which a well-defined decay rate can be associated to the quasiparticles.

\index{Propagator!analytic structure}

\index{Propagator!Feynman}
\index{Propagator!Pauli-Jordan}
\index{Propagator!retarded}
\index{Propagator!advanced}
\index{Propagator!Hadamard}
\index{Propagator!retarded}

When the retarded and advanced propagators can be approximated by
\begin{subequations} \label{UnstableQuasi}
\begin{align}
	\GR(\omega,\vect p) &\approx  \frac{-i Z_\vect p}{-\omega^2 +R^2_{\vect p} - i \omega 
	\Gamma_{\vect p}},\\
	\GA(\omega,\vect p) &\approx  \frac{i Z_\vect p}{-\omega^2 +R^2_{\vect p} + i \omega 
	\Gamma_{\vect p}},
\end{align}
the Pauli-Jordan propagator (or spectral function) is given by
\begin{align}
	G^{(1)}(\omega,\vect p) &\approx  \frac{ 2 |\omega| Z_\vect p \Gamma_\vect p   }
	{\left(-\omega^2+E_\vect p^2\right)^2 + \left( \omega \Gamma_\vect p\right)^2}. 
\end{align}
So far we have analyzed the propagators which are independent of the occupation number. Other propagators depend on the occupation number. 
Approximating $n(|\omega|)$ by the value at the pole, $n_\vect p := n(E_\vect p)$, we obtain from \Eqref{SigmaNCorr}
\begin{equation}\label{SigmaNRPole}
	\SigmaN(E_\vect p,\vect p) \approx  -2i \Im \SigmaR(E_\vect p,\vect p)  (1+2n_\vect p),
\end{equation}
where $n_\vect p$ can be interpreted as the occupation number of the mode with momentum $\vect p$.
Combining the above equation with Eqs.~\eqref{PropsSigma} yields
\begin{align}
	G(\omega,\vect p) &\approx  \frac{ 2 \omega Z_\vect p \Gamma_\vect p (1 + 2n_\vect p)  }
	{\left(-\omega^2+E_\vect p^2\right)^2 + \left( \omega \Gamma_\vect p\right)^2}. 
\end{align}
Finally, the Feynman and Dyson propagators and the Whightman functions can be represented by
\begin{align}
	\GF(\omega,\vect p) &\approx  \frac{-i Z_\vect p (-\omega^2+E_\vect p^2)+ |\omega| Z_\vect p  \Gamma_\vect p \left( 2n_\vect p + 1 \right)  }
	{\left(-\omega^2+E_\vect p^2\right)^2 + \left( \omega \Gamma_\vect p\right)^2}, \\
	\GD(\omega,\vect p) &\approx  \frac{i Z_\vect p (-\omega^2+E_\vect p^2)+ |\omega| Z_\vect p  \Gamma_\vect p \left( 2n_\vect p + 1 \right)  }
	{\left(-\omega^2+E_\vect p^2\right)^2 + \left( \omega \Gamma_\vect p\right)^2},\\
	G_+(\omega,\vect p) &\approx  \frac{ 2|\omega| Z_\vect p  \Gamma_\vect p \left[ n_\vect p + \theta(\omega)\right]  }
	{\left(-\omega^2+E_\vect p^2\right)^2 + \left( \omega \Gamma_\vect p\right)^2}, \label{GplusGuay}\\
	G_-(\omega,\vect p) &\approx  \frac{ 2|\omega| Z_\vect p  \Gamma_\vect p \left[ n_\vect p+ \theta(-\omega)\right]   } 
	{\left(-\omega^2+E_\vect p^2\right)^2 + \left( \omega \Gamma_\vect p\right)^2}.
\end{align}

\end{subequations}

\chapter{Quasiparticles: the real-time approach} \label{IV}

\section{Quantum states corresponding to quasiparticles: the free case} \label{sect:QuasiParticlesFree}

Even in absence of interaction it is not trivial to construct a quantum state which verifies the quasiparticle properties mentioned in the introduction.
As we have seen, in a non-interacting theory in the vacuum the one-particle state is naturally represented by the action of the creation operator on the vacuum, $|\vect p\rangle  = \hat a^\dag_\vect p|0\rangle$, or equivalently by the action of the field operator: $|\vect p\rangle = \sqrt{2E_\vect p} \hat \phi_{-\vect p} |0\rangle$.
\index{Occupation number}
Over a homogeneous and stationary state $\hat\rho$, the positive-energy quasiparticle state can be represented by the action of the creation operator:
\begin{equation}
\begin{split}
	\hat\rho^\pplus _\vect p &:= \frac{1}{n_\vect p + 1}\, 
\hat a^\dag_\vect p \hat\rho \hat a_\vect p 
	 = \frac1{n_\vect p + 1} \sum_{n,\{m\}} \rho_{n,\{m\}} (n+1) |(n+1)_\vect p,\{m\} \rangle\langle (n+1)_\vect p,\{m\}|,
\end{split}
\end{equation}
where the second equality is written in a highly schematic notation in order to avoid cumbersome expressions. Here  
\begin{equation}
	n_{\vect p}=\Tr{(\hat\rho \hat a^\dag_{\vect p} \hat a_\vect p)}
\end{equation}
represents the mean occupation number of the mode with momentum $\vect p$, $|n_\vect p,\{m\} \rangle$ is a basis of the Hilbert space with the $\vect p$ sector singled out, and $\rho_{n,\{m\}}$ is the diagonal representation of the background state in this basis. Notice that the state is normalized: $\Tr{\hat \rho_\vect p^\pplus } = 1$. The factors $n+1$ on the second equality are a consequence of the Bose-Einstein statistics: the probability to create an additional particle increases with the number of already-existing particles. 

The Bose-Einstein statistics are responsible for the following surprising fact: it is a simple exercise to show that the expectation value of the particle number,
\begin{equation}
	\av{\hat N_\vect p}^\pplus = \Tr{\big(\hat\rho^\pplus _\vect p\hat N_\vect p \big)}  = \Tr{\big(\hat\rho^\pplus _\vect p \hat a^\dag_\vect p \hat a_\vect p\big)}  =  \frac{1}{n_\vect p+1} \av{(\hat N_\vect p+1)^2},
\end{equation}
is actually increased more than one with respect to the unperturbed value:
\begin{equation}\label{surprising}
	\av{\hat N_\vect p}^\pplus = n_\vect p + 1 + \frac{\delta n_\vect p^2}{1+n_\vect p} =: n_\vect p + N_\vect p^\pplus,
\end{equation}
where $N^\pplus_\vect p=1 + \delta n_\vect p^2/(1+n_\vect p)$ is the number of excitations and 
\begin{equation}
	\delta n_\vect p^2 = \av{(\hat N_\vect p-n_\vect p)^2} = \av{\hat N^2_\vect p} - n_\vect p^2 \geq 0
\end{equation}
is the dispersion of the number of particles in the background state. For a particle number eigenstate $\delta n_\vect p^2 =0$ and $N^\pplus_\vect p=1$, and for a Gaussian state (see appendix \ref{app:Gaussian}) $\delta n_\vect p^2 = n_\vect p ( n_\vect p +1)$ and $N^\pplus_\vect p=1 + n_\vect p$. 
Therefore, when the background is in a state which is not an eigenstate of the Hamiltonian, $\hat\rho_\vect p^\pplus$ represents slightly more than one additional quasiparticle. The reason for this is purely statistical: the highly occupied components of the state are more likely to become excited, and therefore they tend to gain statistical weight, thereby increasing the particle number. In other words, the action of the creation operator has two simultaneous effects: on the one hand, adding a quasiparticle to the system; on the other hand, increasing the statistical weight of the highly excited states. The statistical contribution is significant when the occupation numbers are large.

The expectation value of the momentum operator is also affected by this statistical effect:
\begin{subequations}
\begin{equation}
	 \av{\hat {\vect P}}^\pplus =  \Tr{\big(\hat\rho^\pplus _\vect p\, \hat{\vect P} \big)}  = \Tr{\big(\hat\rho^\pplus _\vect p\, {\vect p}\, \hat a^\dag_\vect p \hat a_\vect p\big)}  = \vect p N_\vect p^\pplus.
\end{equation}
Anyway the momentum per excitation is $\vect p$. The energy is similarly affected:
\begin{equation}
	\av{\hat H_\text{sys}}^\pplus = \Tr{\big(\hat\rho^\pplus  H_\text{sys} \big)}  =  \Tr{\Big[\hat\rho^\pplus_\vect p \Big( \hat a^\dag_\vect p \hat a_\vect p +a^\dag_{-\vect p} \hat a_{-\vect p}  + 1 \Big)\Big]}  = E^{(0)} + E_\text{QP}  N_\vect p^\pplus,
\end{equation}
\end{subequations}
where $E^{(0)}=\Tr{(\hat\rho \hat H_\text{sys})}$ is the energy of the background and $E_\text{QP} = E_\vect p = \sqrt{\vect p^2+m^2}$ is energy per excitation.

Quasiparticles require small momentum spreads. The spread of the momentum per quasiparticle in the case of Gaussian states is 
\begin{equation}
 	\av{\Delta\vect p}_\text{QP}^\pplus := \frac{1}{N_\vect p^\pplus} \left\{\av{\hat {\vect P}^2}^\pplus - [\av{\hat {\vect P}}^\pplus]^2\right\}^{1/2} = |\vect p|\frac{(4n_\vect p+5n_\vect p^2)^{1/2}}{1+n_\vect p},
\end{equation}
which is a small quantity if the occupation numbers are small. Likewise, the spreads in the particle number and the energy have the same corresponding values. Therefore, in order for the state $\hat\rho_\vect p^\pplus$ to adequately represent a quasiparticle occupation numbers must be small as compared to one.

Since the statistical contribution looks awkward, one can imagine a different way of constructing the quasiparticle excitation:
\begin{equation} \label{AlternativeToSurprise}
	\hat\rho^{\text{(alt)}} _\vect p := \sum_{n,\{m\}} \rho_{n,\{m\}} |(n+1)_\vect p,\{m\} \rangle\langle (n+1)_\vect p,\{m\}|.
\end{equation}
This state has essentially identical properties to $\hat\rho_\vect p^\pplus$, except that it contains exactly one additional particle. Therefore it looks like that it corresponds more closely to the quasiparticle concept we discussed before. However, notice that the above state cannot be easily created from the background via the action of the creation and annihilation operators.  We shall argue in the next section that the state $\hat\rho^\pplus_\vect p$, and not $\hat\rho^{\text{\tiny (alt)}}_\vect p$, appears naturally when studying quasiparticle creation processes.

\index{Quasiparticle!statistical}
We shall consider that the state $\hat\rho^\pplus_\vect p$ represents exactly one additional real quasiparticle of momentum $\vect p$ and energy $E_\vect p$. The additional contribution to the particle expectation number will be called \emph{statistical quasiparticle} contribution. The statistical quasiparticle contribution can be interpreted as being a consequence of the increased knowledge of the background state which is derived from the creation of a real quasiparticle. 

Another novelty is that negative energy excitations, or holes, can also be defined, represented by the action of the annihilation operator:
\begin{equation}\label{holeState}
\begin{split}
	\hat\rho^\pminus _\vect p &:= \frac{1}{n_\vect p}\, 
\hat a_{-\vect p} \hat\rho \hat a^\dag_{-\vect p} = \frac1{n_\vect p} \sum_{n,\{m\}} \rho_{n,\{m\}} n |(n-1)_{-\vect p},\{m\}\rangle\langle (n-1)_{-\vect p},\{m\}|.
\end{split}
\end{equation}
This hole state is also affected by the statistical considerations described above, as it is manifest by showing the expectation value of the number operator:
\begin{equation}
\av{\hat N_\vect p} = \Tr{(\hat\rho^\pminus _\vect p\hat N )}   = n_\vect p - 1 + \frac{\delta n_\vect p^2}{n_\vect p} =: n_\vect p - N_\vect p^\pminus ,
\end{equation} 
where $N^\pminus_\vect p = 1 - {\delta n_\vect p^2}/{n_\vect p}$ is the expected number of negative-energy excitations. Therefore the hole state contains at most one negative-energy excitation. This state has momentum $\vect p$, $\av{\hat{\vect P}}^\pminus  = \vect p N_\vect p$, and has an additional amount of negative energy $E_\vect p$, $\av{\hat{\vect P}}^\pminus  = E^{(0)} - E_\vect p N^\pminus_\vect p$.

For a particle number eigenstate $N_\vect p^\pminus = 1$, but a  is that for a Gaussian state the number of negative energy excitations is actually negative $N_\vect p^\pminus = -n_\vect p$. The reason for this last surprising fact is that the annihilation operator, besides explicitly removing one particle from the state, also enhances the probability of the highly populated sectors of the state, and the last effect is actually dominating. 
We will consider that $\hat\rho_\vect p^\pminus$ represents a single real hole plus some contribution of statistical quasiparticles of opposite momentum.   Again, the statistical quasiparticle contribution can be thought as a consequence of the increased knowledge of the background state coming from the absorption of a quasiparticle (or creation of a hole). In any case, the fact that for Gaussian states the statistical contribution always dominates is an indication that holes cannot be considered true quasiparticles in bosonic systems.
A confirmation of this fact is given by the spread in the momentum per quasiparticle, 
\begin{equation}
 		\av{\Delta\vect p}^\pminus_\text{QP} := \frac1{|N_\vect p^\pminus|} \left\{\av{\hat {\vect P}^2}^\pminus - [\av{\hat {\vect P}}^\pminus]^2 \right\}^{1/2}= |\vect p|(5+1/n_\vect p)^{1/2},
\end{equation}
which is of the order of the expectation value of the momentum per quasiparticle, even if the occupation numbers are small. Even if holes are not true quasiparticles in bosonic systems, it will be important to remember that negative energy excitations are possible when the states are different than the vacuum.

So far we have seen that quasiparticles and holes are respectively created by the creation and annihilation operators. The field operator $\hat\phi_{\vect p}$, being a linear combination of creation and annihilation operators, creates a coherent superposition of quasiparticles and holes. In effect, the state $\hat\phi_{-\vect p} \hat\rho \hat\phi_{\vect p}$ corresponds to the linear superposition of a quasiparticle and a  hole:
\begin{equation}
	\hat\phi_{-\vect p} \hat\rho \hat\phi_{\vect p} = \frac{1}{2E_{\vect k}} \left(\hat a^\dag_{\vect p} + \hat a_{-\vect p}\right) \hat \rho  \left(\hat a_{\vect p} + \hat a^\dag_{-\vect p}\right).
\end{equation}
For states characterized by low  occupation numbers the quasiparticle contribution dominates over the hole contribution.

\index{Particle detector}

\section{Creation of quasiparticles}\label{sect:QuasiCreation}

Let us argue that only the states created with the creation and annihilation operators, or equivalently with the field operator, are physically meaningful. 
In a realistic situation quasiparticles are created with the interaction of the field with some external agent. This external agent can be modeled by a ``quasiparticle emitter''---which  essentially coincides with the usual particle detector considered in the analysis of the Unruh effect \cite{Unruh76,BirrellDavies}: an external device described by some quantum mechanical degree of freedom $Q$ (which can correspond for instance to a harmonic oscillator or a two-level system). The device starts in an excited state. For simplicity, let us assume that the emitter is linearly coupled with one pair of field modes, $g_Q Q(t) [\phi_\vect p(t)+\phi_\vect{-p}(t)] $.\footnote{Notice that in this simplified model the emitter cannot couple to a single mode of the field because that couping would not be momentum-conserving (or, from another point of view, would not be Hermitian).}  The coupling constant $g_Q$ is very small so that the external device, other than emitting quasiparticles, does not significantly perturb the dynamics of the field. The initial state of the field plus detector is assumed to be $\hat\rho \otimes |1_Q\rangle \langle 1_Q|$. The aim is to find the final state for the field $\hat\rho'$ when the detector is measured in its unexcited state $|0\rangle$. 

The time evolution of the entire system under the interaction is given by
\begin{equation}\label{TotalEvolution}
	\hat\rho_\text{total}(t)=
	T \expp{-i\int_{t_0}^t g_Q \hat\phi_\text{I}(s) \hat Q_\text{I}(s) \vd s }
	\hat\rho \otimes |1_Q\rangle \langle 1_Q|
	T \expp{i\int_{t_0}^t g_Q \hat\phi_\text{I}(s) \hat Q_\text{I}(s) \vd s },
\end{equation}
where the subindex $\text I$ indicates interaction picture, and $\phi_\text I := \phi_\vect p+ \phi_{-\vect{p}}$. Since the coupling is small, the above equation can be expanded as
\begin{equation*}
	\hat\rho_\text{total}(t)=
	\left[1-i\int_{t_0}^t g_Q \hat\phi_\text{I}(s) \hat Q_\text{I}(s) \vd s \right]
	\hat\rho \otimes |1_Q\rangle \langle 1_Q|
	\left[1+{i\int_{t_0}^t g_Q \hat \phi_\text{I}(s) \hat Q_\text{I}(s) \vd s }\right],
\end{equation*}
or, developing the expression
\begin{equation} \label{76}
\begin{split}
	\hat\rho_\text{total}(t)&=
	\hat\rho \otimes |1_Q\rangle \langle 1_Q|
	- i g_Q \int_{t_0}^t \ud s g_Q \big[\hat\phi_\text{I}(s),\hat\rho\big] \otimes \big[ \hat Q_\text{I}(s) ,  |1_Q\rangle \langle 1_Q| \big]\\
	&\quad + g_Q^2 \int_{t_0}^t \ud s \int_{t_0}^t \ud {s'} \hat\phi_\text{I}(s) \hat\rho\, \hat\phi_\text{I}(s') \otimes
	\hat Q_\text{I}(s)  |1_Q\rangle \langle 1_Q| \hat Q_\text{I}(s').
\end{split}
\end{equation}
The final state for the system, if the emitter is found unexcited at time $t$, is given by the projection into the ground state of the emitter \cite{GalindoPascual}:
\begin{equation}
	\hat\rho' = \frac{ \Tr_\text{emitter} \left( \hat\rho_\text{total}(t) |0_Q\rangle \langle0_Q| \right)}{ \Tr_\text{total} \left( \hat\rho_\text{total}(t) |0_Q\rangle \langle 0_Q| \right)}.
\end{equation}
Developing \Eqref{76} we find
\begin{equation*}
\begin{split}
	\hat\rho' &= N  \int_{t_0}^t \ud s \int_{t_0}^t \ud {s'} \hat\phi_\text{I}(s) \hat\rho\, \hat\phi_\text{I}(s') \otimes 
	\langle 0 | \hat Q_\text{I}(s)  |1_Q\rangle \langle 1_Q| \hat Q_\text{I}(s') | 0 \rangle \\
	&= N'  \int_{t_0}^t \ud s \int_{t_0}^t \ud {s'} \expp{i \Omega(s-s')} \hat\phi_\text{I}(s) \hat\rho\, \hat\phi_\text{I}(s'),
\end{split}
\end{equation*}
where $\Omega$ is the frequency of the harmonic oscillator and $N$ and $N'$ are normalization constants. Expanding in terms of creation and annihilation operators we get:
\begin{equation*}
\begin{split}
	\hat\rho' &= N'' \int_{t_0}^t \ud s \int_{t_0}^t \ud {s'} \expp{i \Omega(s-s')}\Big(  \hat a_\vect p^\dag \hat\rho\, \hat a_\vect p \expp{-iE_\vect p(s-s')}+\hat a^\dag_\vect p \hat\rho\,\hat a_{-\vect p}^\dag \expp{-iE_\vect p(s+s')} 
	\\ &\qquad +  \hat a_{-\vect p} \hat\rho\, \hat a_\vect p \expp{iE_\vect p(s+s')} +  \hat a_{-\vect p} \hat\rho\, \hat a^\dag_{-\vect p} \expp{iE_\vect p(s-s')} 
	+ \hat a_{-\vect p}^\dag \hat\rho\, \hat a_{-\vect p} \expp{-iE_\vect p(s-s')}\\ &\qquad +\hat a^\dag_{-\vect p} \hat\rho\,\hat a_{\vect p}^\dag \expp{-iE_\vect p(s+s')} 
	 +  \hat a_{\vect p} \hat\rho\, \hat a_{-\vect p} \expp{iE_\vect p(s+s')} +  \hat a_{\vect p} \hat\rho\, \hat a^\dag_{\vect p} \expp{iE_\vect p(s-s')}\Big).
\end{split}
\end{equation*}
Let us assume that the frequency of the oscillator is tuned so that $\Omega= E_\vect p$. In this case, for sufficiently large time lapses the dominant value of the integral is given by the stationary value of the integral of the first term, which amounts to considering energy conservation. In that case
\begin{equation} \label{physicalStateQP}
\begin{split}
	\hat\rho' &\approx N''' \big(\hat a_\vect p^\dag \hat\rho\,\hat a_\vect p + \hat a_{-\vect p}^\dag \hat\rho\,\hat a_{-\vect p} \big) = \frac12 \big(\hat\rho_\vect p^\pplus +\hat\rho_{-\vect p}^\pplus\big).
\end{split}
\end{equation}
Therefore, upon deexcitation of the emitter, the system gets promoted to a superposition of two quasiparticle states. The argument could be repeated with the same measuring device in the ground state, now  interpreted as a particle detector. When the measuring device gets excited, the state of the field colapses to a superposition of the hole states $\hat\rho^\pminus_\vect p$ and $\hat\rho^\pminus_{-\vect p}$.

\section{Quantum states corresponding to quasiparticles: the interacting case}

\index{Quasiparticle!quantum state}

\label{Asymptotic quasiparticles}
We have just seen that when no interaction is present, positive-energy quasiparticles are created by the action of the creation operator, and negative-energy holes are created by the annihilation operator. If interaction is present and quasiparticles are stable, similarly to the vacuum case  (see section \ref{sect:Asympt}), one can think of defining asymptotic quasiparticle fields and states, from which the two-point correlation functions can be reproduced. Namely, quasiparticle states correspond to
\begin{subequations} \label{ThermalAsympt}
\begin{equation}
	\hat\rho^\pplus_\vect p \cong \frac{1}{\bar n_\vect p + 1}\, 
\hat \ar^\dag_\vect p \hat\rho \hat \ar_\vect p ,
\end{equation}
and negative energy holes correspond to
\begin{equation} 
	\hat\rho^\pminus _\vect p \cong \frac{1}{\bar n_\vect p}\, 
\hat \ar_{-\vect p} \hat\rho \hat \ar^\dag_{-\vect p},
\end{equation}
\end{subequations}
where $\bar n_{\vect p}=\Tr{(\hat\rho \hat \ar^\dag_{\vect p} \hat \ar_\vect p)}$ is the expectation value of the number of asymptotic quasiparticles in the state with momentum $\vect p$, and we recall the symbol $\cong$ means equivalence when evaluated in matrix elements in the asymptotic limit.  The asymptotic creation and annihilation operators $\hat\ar$ and $\hat\ar^\dag$ are related to the asymptotic field operator through
\begin{equation}\label{AsymptFieldOp}
	\hat\phir_\vect p = \frac{1}{\sqrt{2 E_\vect p}} \big( \hat \ar _\vect p + \hat \ar^\dag_{-\vect p}\big),
\end{equation}
where $E_\vect p$ is the quasiparticle energy. The asymptotic field, which obeys free equations of motion, $\ddot \phir + (m^2+ E_\vect p^2) \phir = 0$, is connected to the usual field through
\begin{equation}
	\hat\phir_\vect p \cong Z_\vect p^{-1/2} \hat\phi_\vect p,
\end{equation}
where $Z_\vect p$ is defined in \Eqref{StableQuasiparticle}. See Refs.~\cite{NarnhoferEtAl83,GreinerLeupold98} for comparable approaches.

However, since quasiparticles are generically unstable and therefore not completely well-defined from a strict point of view, it is not worth pursuing a very formal description. Therefore we will blurry the distinction between the usual and the asymptotic fields, and work with the usual field operator but rescaled a factor  $Z_\vect p^{-1/2}$, \ie, we will set $\mathcal Z_\vect p = Z_\vect p$ [see Eqs.~\eqref{OQSAction}]. With this assumption  equations \eqref{ThermalAsympt} can be more simply restated:
\begin{equation}
	\hat\rho^\pplus_\vect p \approx \frac{1}{n_\vect p + 1}\, 
\hat a^\dag_\vect p \hat\rho \hat a_\vect p,  \qquad
	\hat\rho^\pminus _\vect p \approx \frac{1}{n_\vect p}\, 
\hat a_{-\vect p} \hat\rho \hat a^\dag_{-\vect p}. 
\end{equation}
Bear in mind that this representation is only valid when studying (approximately) asymptotic properties.

When we presented the open quantum system analysis of the field modes (see the introduction), we also introduced two free parameters, $E_\vect p$ and $\mathcal Z_\vect p$. At this point we have, on the one hand, a criterion to fix their value, and, on the other hand, a physical interpretation for both of them. With respect to the former, $E_\vect p$ must be fixed to the value of the real part of the pole of the propagator, and represents the physical energy of the quasiparticle excitation. With respect to the latter, $\mathcal Z_\vect p$ measures the probability that the interacting field operator excites the quasiparticle state. Roughly speaking, by rescaling the field a factor $\mathcal Z_\vect p=Z_\vect p$ we  ensure that creation and annihilation operators create and destroy quasiparticles with the proper normalization. From now on we shall assume that such a scaling has been performed so that $\mathcal Z_\vect p=Z_\vect p$. From a practical point of view this amounts to setting $Z_\vect p=1$ in equations \eqref{UnstableQuasi}.

\section{Time evolution of the quasiparticle excitations}\label{sect:Quasitime}

The aim is to find the time evolution of the expectation value of the Hamiltonian of the system. We shall make use of the open quantum system viewpoint presented in the introduction. Let us start with the positive energy excitations, and focus on the state with momentum $\vect p$. The expectation value of the Hamiltonian operator in such a quantum system is given by
\begin{equation}
	E^\pplus(t,t_{0};\vect p) := \av{\hat H_\text{sys} (t)}_\pplus  = \frac{1}{ n_\vect p + 1} \Tr{ \big(\hat a_\vect p^\dag\hat\rho\hat a_\vect p U(t_0,t)\hat H_\text {sys} U(t,t_{0})   \big) },
\end{equation}
where the system Hamiltonian is given by \Eqref{HSys}.
Discarding the connected part of the 4-point correlation functions, as it is manifest in the following application of the Wick theorem [see appendix \ref{app:Gaussian} and in particular \Eqref{WickMixed}]:
\begin{equation*}
\begin{split}
	E^\pplus(t,t_{0};\vect p) &\approx \frac{1}{n_\vect p + 1} \Big\{
	 \Tr_\text{sys}  \big( \hat \rho_\text{s} \hat a_\vect p \hat a_\vect p^\dag ) \Tr_\text{sys}  \big( \hat \rho_\text{s} \hat H_\text{sys} \big)    \\
	&\quad+E_\vect p \Tr_\text{sys}  \big[ \hat \rho_\text{s} \, \hat a_\vect p U(t_0,t) \hat a_\vect p^\dag U(t,t_0) \big] 
	\Tr_\text{sys}  \big[ \hat \rho_\text{s} \, \hat a_\vect p^\dag U(t,t_0) \hat a_\vect p U(t_0,t) \big] \Big\},
\end{split}
\end{equation*}
which can be rewritten as
\begin{equation*}
	E^\pplus(t,t_{0};\vect p) \approx E^{(0)} +\frac{E_\vect p}{n_\vect p + 1}
	\left| \Tr_\text{sys}  \big[ \hat \rho_\text{s} \, \hat a_\vect p U(t_0,t) \hat a_\vect p^\dag U(t,t_0) \big]\right|^2 ,
\end{equation*}
where $E^{(0)}:=\Tr_\text{sys}  \big( \hat \rho_\text{s} \hat H_\text{sys} \big)=E_\vect p ( n_\vect p + n_{-\vect p}+1/2 )$ is the energy of the reduced subsystem before the perturbation, or as
\begin{equation}
	E^\pplus(t,t_{0};\vect p) \approx E^{(0)} +  \frac{E_\vect p}{n_\vect p + 1}
	\left|\frac{1}{2E_\vect p}(E_\vect p + i\partial_t)(E_\vect p -i \partial_{t_0}) G_+(t,t_0;\vect p)\right|^2,
\end{equation}
where we have used the expression of the creation and annihilation operators in terms of the asymptotic field and asymptotic canonical momentum, \Eqref{46}, and the ``field'' representation of the canonical momentum, $\hat\pi_\vect p (t) = \dot{\hat\phi}_\vect p(t)$.

Using translation invariance the energy of the perturbation can be rewritten as
\begin{subequations}
\begin{equation} \label{EnergyGuay}
	E^\pplus(t,t_{0};\vect p) \approx E^{(0)} + \frac{E_\vect p}{n_\vect p + 1}
	\left|I(t,t_0;\vect p)\right|^2 .
\end{equation}
where
\begin{equation}  \label{IntegralI}
	I(t,t_0;\vect p)=\frac{1}{2E_\vect p}(E_\vect p + i\partial_{t})^2  G_+(t,t_0;\vect p)
\end{equation}
Using \Eqref{GplusGuay}, the expression $I(t,t_0;\vect p)$ can be written as:
\begin{equation} \label{IntegralI2}
 I(t,t_0;\vect p) =  \int \frac{\vd \omega}{2\pi} \expp{-i\omega(t-t_0)} \frac{ |\omega| (\omega+ E_\vect p)^2  \Gamma_\vect p \left[ n_\vect p + \theta(\omega)\right]  }{\left(-\omega^2+E_\vect p^2\right)^2 + \left( \omega \Gamma_\vect p\right)^2}.
\end{equation}
\end{subequations}
The above integral is evaluated by integration in the complex plane in appendix \ref{app:Integral}. Neglecting $\Gamma_\vect p$ in front of $E_\vect p$ it is found $ I(t,t_0;\vect p) \approx E_\vect p ( n_\vect p+1) \expp{-\Gamma_\vect p (t-t_0)}$, and therefore the result is
\begin{equation}\label{TimeEvolQuasi1}
	E^\pplus(t,t_{0};\vect p) \approx E^{(0)}+ E_\vect p (n_\vect p +1) \expp{-\Gamma_\vect p (t-t_0)}.
\end{equation}
The factor $n_\vect p +1$ is the statistical factor (recall that for a Gaussian state the expected number of excitations is $N_\vect p^\pplus =  n_\vect p +1$). Discounting this statistical factor the energy of the quasiparticle thus evolves as:
\begin{equation}\label{TimeEvolQuasi2}
	E^\pplus_\text{QP}(t,t_{0};\vect p) \approx  E_\vect p  \expp{-\Gamma_\vect p (t-t_0)}.
\end{equation}

For the hole excitations we find a similar result
\begin{equation}
	E^\pminus(t,t_{0};\vect p) \approx E^{(0)} + E_\vect p n_\vect p \expp{-\Gamma_\vect p (t-t_0)}.
\end{equation}
Recall that, because of statistical effects, for Gaussian states the initial expectation value for the energy of the hole excitations is given by $E_\vect p n_\vect p$.

\index{Decay rate}
Equations \eqref{TimeEvolQuasi1} and \eqref{TimeEvolQuasi2}  are the main results of this section. From these equations $E_\vect p$ can be identified as the energy of the quasiparticles, in agreement to the results of the previous section, and $\Gamma_\vect p$ as the decay rate of the quasiparticles, according to the interpretation of the imaginary part of the self-energy as the decay rate---recall that $E^2_\vect p$ and $E_\vect p\Gamma_\vect p$ correspond to the real and imaginary parts of the self-energy respectively.

Notice that four basic assumptions have been used in order to reach these results. First, we have introduced an approximately asymptotic representation for the field operators in terms of the creation and annihilation operators, which is valid if we are considering the evolution over times much longer than the characteristic interaction time. Second, and related to this, we have used a near-on shell approximation for the propagators, which is also correct if we are investigating long times (but not extremely long, as discussed in section \ref{sect:TimeEvol}). Third, we introduced a Gaussian approximation, which makes the problem solvable in term of the two-point correlation functions. Finally, we have also assumed that the retarded propagator has the analytic structure given by \Eqref{AnalyticStandard}. As we have already mentioned, in presence of massless excitations there can be situations in which the analytic structure is different, leading to non-exponential modifications of the decay law \eqref{TimeEvolQuasi1}, as in the case of hot QED plasmas \cite{BlaizotIancu96,BlaizotIancu97,BoyanovskyDeVega98,BoyanovskyDeVega01,BoyanovskyEtAl98}.

It could be argued that the derivation presented in this section is somewhat cyclic, because we are assuming from the beginning that $E_\vect p$ is the frequency corresponding to the 2-mode system. Following what is done in Ref.~\cite{ArteagaThesis}, one could instead assume that the frequency of the system has an unknown value $E_\vect p'$. We would then find a rapid oscillatory behavior for the expectation value of the energy. Only when $E_\vect p'=E_\vect p$ the energy follows a smooth exponential decay as is expected on physical grounds.

We have computed the time evolution of the propagators [and hence the time evolution of the energy, according to Eqs.~\eqref{EnergyGuay} and \eqref{IntegralI}] by going to the frequency space and taking profit of the spectral analysis of the propagators carried out in the previous section. Alternatively, it is possible to bypass the spectral analysis by considering the equation of motion followed by the propagators, which in general is an integro-differential equation. Making a local approximation to the self-energy in terms of the energy an the decay rate, the equation of motion turns into an ordinary differential equation which can readly be solved \cite{Arteaga08b}. Both methods are physically equivalent and lead to identical results.

\chapter{Alternative mean-field-based approaches}\label{V}

\index{Linear response}
\section{Linear response theory}

The standard way of analyzing the quasiparticle properties is with the aid of the linear response theory \cite{Kapusta,LeBellac,FetterWalecka,Reichl}. In linear response theory the Hamiltonian $H$ of a quantum mechanical system is perturbed with some external perturbation $V(t)$ at time $t_0$, so that the total Hamiltonian is $H + V(t)$ for $t>t_0$. Given some quantum operator $\hat O(t)$, it is a simple exercise to show that to first order in the potential the expectation value of that operator in is given by
\begin{equation}\label{313}
	\av{\hat O(t)} = \av{\hat O(t)}_0 + \delta\av{\hat O(t)} =  \av{\hat O(t)}_0  - i  \int_{t_0}^\infty \Tr{ \big( \hat \rho\, \theta(t-t') \big[\hat O_\text{I}(t),\hat V_\text{I}(s)\big] \big) },
\end{equation}
where I indicates interaction picture with respect to the external perturbation and $\av{\cdot}_0$ indicates the expectation value in absence of the external potential.  

Linear response theory is usually applied to a free or interacting scalar field theory, with the following identifications:
\begin{equation*}
	\hat V(t) = \int \ud[3]{\vect x} j(t,\vect x) \hat\phi(\vect x),
	\qquad
	\hat O(t) = \hat\phi(\vect x).
\end{equation*}
Then, from \Eqref{313}, one obtains:
\begin{equation}
	 \delta\av{\hat\phi(t,\vect x)} = -i \int \ud[4]{x} \GR(x,x') j(x').
\end{equation}
For the case of an impulsive perturbation, $j(x) = j(\vect q) \expp{i\vect q \cdot \vect x} \delta(t)$, if the  retarded propagator is approximated as
\begin{equation*}
	\GR(\omega,\vect p) \approx \frac{  i Z_\vect p }{(2E_\vect p)(\omega -E_\vect p + i \Gamma_\vect p/2)}, \quad
	\omega \sim E_\vect p,
\end{equation*}
the following result for the dynamics of the expectation value of the scalar field is obtained \cite{LeBellac}:
\begin{equation} \label{LinearResponseNaive}
	\delta\av{\hat\phi(t,\vect x)} \approx - 
	\frac{i j(\vect q)}{2 E_\vect p} \theta(t) \expp{ i(\vect q \cdot \vect x - E_\vect p t)} \expp{-\Gamma t/2}.	
\end{equation}
The fluctuation oscillates with a frequency $E_\vect p$ and decays with a rate $\Gamma_\vect p/2$. Those are interpreted as the energy and damping rates of the excitations. Notice that the energy of the oscillation decays with a rate $\Gamma_\vect p$.

In non-relativistic $N$-body theory, linear response is usually applied to the fermion density, instead of the field itself, and the final result depends on the density correlation functions \cite{FetterWalecka}.  Beyond the  field theory context, linear response has many different applications in statistical mechanics \cite{Reichl}, solid state physics \cite{AshcroftMermin}, and even gravitation \cite{AndersonEtAl03}.

\section{Effective dynamics of the mean field}


Another possible method to study the quasiparticle properties has been the analysis of the effective dynamics of the mean field. This method has been previously used  in the literature by Weldon \cite{Weldon98} and by Drummond and Hathrell \cite{DrummondHathrell80} in a curved background context (see also Ref.~\cite{Shore03b}). Let us briefly review it in the CTP context.
 
Functionally
differentiating the CTP effective action we get the effective
equations of motion for the expectation value of the field,
$\varphi := \Tr (\hat\rho\hat\phi)$ (see  appendix \ref{app:CTP}):
\begin{equation}\label{CTPEqsMot}
    \left. \derf{\Gamma[\varphi_1,\varphi_2]}{\varphi_{1}(x)}
    \right|_{\varphi_1=\varphi_2=\varphi} = 0.
\end{equation}
The effective action can always be expanded in the proper vertices
$\Gamma^{a_1\cdots a_r} (x_1,\ldots,x_r)$:
\begin{equation}
\begin{split}
    \Gamma[\varphi_1,\varphi_2] = \sum_r \frac{1}{r!}\int
    &\  \ud[4]{x_1} \cdots  \vd[4]{x_r}\Gamma^{a_1\cdots a_r} (x_1,\ldots,x_r)
    \varphi_{a_1}(x_1)\cdots
    \varphi_{a_r}(x_r).
\end{split}
\end{equation}
 A straightforward generalization of the usual
argument (see \eg\ Ref.~\cite{Peskin}) shows that this 2-point vertex
corresponds to the inverse propagator,
\begin{equation}
    \Gamma^{ab}(x,y) = i (G^{-1})^{ab}(x,y).
\end{equation}
The Schwinger-Dyson equation, which defines the self-energy $\Sigma^{ab}(x,x')$,
\begin{equation*}
    G_{ab}(x,y) = G_{ab}^{(0)}(x,y) - i \int 
    \ud[4]{z}  \ud[4]{w}  G_{ac}^{(0)}(x,z) \Sigma^{cd}(z,w)
    G_{cb}(w,y),
\end{equation*}
can be manipulated to give
\begin{equation}
    (G^{-1})^{ab}(x,y) = A^{ab}(x,y) +  i\Sigma^{ab}(x,y),
\end{equation}
where $A^{ab}(x,y)$ is the inverse of the free propagator,
\begin{equation} \label{InverseProp}
    A^{ab}(x,y)  =  [(G^{(0)})^{-1}]^{ab}(x,y) = c^{ab}
       i (-\Box^2+m^2) \delta^{(4)}(x-y),
\end{equation}
with $c^{ab}=\diag(1,-1)$. We see that the 2-point vertex can be
expressed as
\begin{equation} \label{2VertexSigma}
    \Gamma^{ab}(x,y) = i A^{ab}(x,y) - \Sigma^{ab}(x,y).
\end{equation}
Hence the 2-point vertex essentially corresponds to the
self-energy. Other proper vertices also have
similar interpretations in terms of one-particle irreducible diagrams.

If the relevant vertex is the 2-particle
vertex $\Gamma^{ab}$, the effective equations of motion can be expressed as
\begin{equation}
\begin{split}
        \left. \derf{\Gamma}{\varphi_{1}}
    \right|_{\varphi_1=\varphi_2=\varphi} &= \int \vd[4]{y} \,[\Gamma^{11}(x,y)+
    \Gamma^{12}(x,y)] \varphi(y) = 0   ,
\end{split}
\end{equation}
which, taking into account Eqs.~\eqref{InverseProp} and
\eqref{2VertexSigma} can be expanded as
\begin{equation}\label{EqMotion}
       (-\Box_x+m^2)\varphi (x)+
     \int \vd[4]{y}  \,
    \Sigma_\mathrm R(x,y) \varphi(y)  = 0,
\end{equation}
where $\Sigma_\mathrm R(x,y) = \Sigma^{11}(x,y) +
\Sigma^{12}(x,y)$ is the retarded self-energy. 
Introducing the Fourier transform around the mid point, the above equation can be rewritten as
\begin{equation} \label{DispRelNE}
    p^2+m^2+ \Sigma_\mathrm R(p;X) = 0.
\end{equation}
Eq.~\eqref{DispRelNE} amounts to
finding the poles of the retarded propagator. Thus, in flat spacetime the self-energy and effective action methods lead to the same dispersion relation
provided we use a CTP approach in both situations and we neglect
vertices with three external particles or more. However, let us see that the interpretation of the dispersion relation is slightly different. Provided we can approximate,
\begin{equation*}
	p^2+m^2+ \Sigma_\mathrm R(p;X) \approx -\omega^2 + m^2+ \vect p^2 + E_\vect p - i\omega\Gamma_\vect p
\end{equation*}
the solution to $\eqref{EqMotion}$ can be written as
\begin{equation}
	\varphi(x) \approx \int \udpi[3]{\vect k} \left [ A(\vect p)
	\expp{-i\vect k \cdot \vect x} 
	+ B(\vect p)
	\expp{i\vect k \cdot \vect x} \right] \expp{-i E_\vect p t} \expp{-\Gamma_\vect p t/2}.
\end{equation}
Therefore the real and imaginary parts of the poles have the role of the frequency of the damping rate of the mean field excitation.

\section{The mean-field and quasiparticle methods compared}

The mean-field-based approaches are simpler than the methods that we have developed in this paper, highlight the importance of the retarded propagator, and quickly relate the real and imaginary parts of the poles of the propagator to the energy and the decay rate respectively. However, the expectation value of the field operator is not a usual observable in field theory. Moreover, the kind of perturbations considered does not correspond to quasiparticles, since the latter, which are of the form $\hat a^\dag \hat\rho \hat a$, have vanishing expectation values for the field operator. Given all that, we believe that in relativistic field theory a linear response-based approach based on the study of the dynamics of expectation value of the field operator is not  well-suited to study the dynamical properties of quasiparticles in the regime of validity of the quasiparticle description. It is adequate though to describe the hydrodynamic regime.

Relativistic field theory in the hydrodynamic limit \cite{Forster,Groot} can be understood as an effective theory describing the dynamics of long wavelength and short frequency perturbations \cite{SonStarinets07}. It has been extensively studied in the literature: for instance, let us mention that the evaluation of the hydrodynamic transport coefficients has been carried out by Jeon and Yaffe \cite{Jeon95,JeonYaffe96} and by Calzetta \emph{et al.} \cite{CalzettaEtAl00}, and that Aarts and Berges have studied the non-equilibrium time evolution of the spectral function in this regime \cite{AartsBerges01}.

The limit of long wavelengths and short frequencies  corresponds to considering small momenta. For Gaussian systems other than the vacuum, small-momentum modes have occupation numbers of the order of one or larger, an therefore this regime is not adequately described by a quasiparticle description, as commented in the introduction. Therefore, the hydrodynamic and quasiparticle descriptions  represent two different complementary descriptions valid in two different regimes. The classical-like perturbations at low momenta are adequately described by a hydrodynamic description, and the individual particle-like perturbations at high momenta can be analyzed within a quasiparticle formalism. 

In any case, we have seen that both the  dynamics of the mean field and the quasiparticles can be characterized with the retarded self-energy. This can be understood in several ways, one of them being the following: loosely speaking, when the Gaussian approximation is introduced and higher order correlation functions are neglected, the system behaves as if it were effectively linear, and thus the dynamics of all relevant quantities is essentially determined by the solution of a corresponding stochastic problem \cite{GardinerZoller,GreinerLeupold98,CalzettaRouraVerdaguer03}. In this context it is not surprising that the elementary dynamics of both the mean field and the quasiparticles can be described with the same elements.

\chapter{Summary and discussion} \label{VI}

\enlargethispage{\baselineskip}

In this paper we have presented two different approaches to the analysis of  the quasiparticle properties in relativistic field theory: first, a frequency-based approach, in which the properties of the excitations have been deduced from the analysis of the spectral representation and, second, a real-time approach, wherein the quantum states corresponding to the quasiparticle excitations have been explicitly constructed, and the time evolution of the expectation value of the energy in those states has been studied. In both approaches it has been possible to show that the real and imaginary part of the self-energy determine the energy and the decay rate of the quasiparticles respectively---see Eqs.~\eqref{RGammaBasic}.  Although previous evidences existed, to our knowledge this is the first systematic corroboration that the real part of the self-energy determines the physical energy of quasiparticles in non-vacuum relativistic field theory. The dynamics of the quasiparticle excitation can be also encoded in the form of a generalized dispersion relation [see Eq.~\eqref{GenDispRel}]:
\begin{equation}\label{generalizedDispRel}
	\mathcal E^2 = E_\vect p^2 - i E_\vect p \Gamma_\vect p =  m^2 + \vect p^2 + \SigmaR(E_\vect p,\vect p),
\end{equation}
which needs not be Lorentz-invariant. 

Several additional interesting points have been illustrated by the real-time approach. We have explored the open quantum system viewpoint for the quantum field modes, and have found a physical criterion to fix the values of the system frequency and field renormalization parameters, equivalent to the on-shell renormalization condition in the vacuum. Using a very simplified model of a quasiparticle creation device, we have discussed that quasiparticles are adequately described by the action of the creation operator on the background state for long observation times---see \Eqref{physicalStateQP}. At the same time, we have shown that the quantum state corresponding to the quasiparticle contains actually more than one particle excitation. We have argued that this result is actually a statistical effect, due to the fact that sectors with higher occupation numbers are more likely to be excited, because of the Bose-Einstein statistics. We have also built the hole state [\Eqref{holeState}], but have discussed that it cannot be properly considered a quasiparticle state for bosonic systems.

Finally, by comparing our results with other mean-field-based approaches, we have argued that the latter are more suited for the study of classical-like configurations in the hydrodynamic limit rather than to the study of the properties of individual quasiparticles. 

\index{Dispersion relation}
\index{Mass!effective}
\index{Mass!thermal}



Most expressions in this paper have referred to homogeneous, isotropic and stationary backgrounds. In fact, the main results depend, first, on the existence of the diagonal relation \eqref{GSigmaDiagonal} between the self-energy and the propagator in Fourier space (which demands homogeneity and stationarity); second, on the diagonalization of the density matrix in the basis of eigenstates of the Hamiltonian (which is implied by stationarity), and, third, on the equivalence of the $\vect p$ and $-\vect p$ modes (which requires isotropy). Therefore, it appears that the quasiparticle interpretation will be completely spoiled when the background becomes non-stationary, non-homogeneous or non-isotropic. 
Let us argue that this is not the case, and that a particle interpretation is still feasible for non-homogeneous and non-stationary backgrounds provided that the characteristic scales of variation of the background state are sufficiently large. In the following we shall assume that $L$ is the typical length scale in which the background changes significantly, and $T$ is the typical timescale of evolution of the background. 

When the backgrounds are non-homogeneous or non-stationary there is still an approximate diagonal relation between the retarded propagator and retarded self-energy [Eq.~\eqref{GSigmaDiagonal}], provided $ L \gg 1/E$ and $T \gg 1/E$, where $E$ is the typical energy involved. The retarded propagator will have the usual analytic structure, with the only difference that now the location of the poles will be time and space dependent, and therefore the generalized dispersion relation will also be time and space dependent:
\begin{equation}
	\mathcal E^2 = m^2 + \vect p^2 + \SigmaR\boldsymbol(E_\vect p(t,\vect x),\vect p;t,\vect x\boldsymbol) = 0.
\end{equation}
On the other hand, for timescales much smaller than $T$, the background can be considered stationary. Hence, if one considers energies which correspond to those short timescales, $E \gg 1/T$, the density matrix components corresponding to those energies are diagonal in the basis of eigenstates of the Hamiltonian. Therefore the properties which depend on the diagonalization, such as the spectral representation, are still valid.
Finally, if the origin of the anisotropy is the inhomogeneity of the state, for the relevant energy scales the anisotropy will also be negligible if the condition $L \gg 1/E$ is fulfilled. However, it is  possible that the system is homogeneous but anisotropic, and that the magnitude of the anisotropy is large. In this case some of the expressions we have written down will no longer be quantitatively valid, since we have demanded the equivalence of the forward and backward modes for a given momentum. However in Refs.~\cite{ArteagaParentaniVerdaguer05,ArteagaThesis} it is shown with a particular example that the general picture is not essentially modified.

\enlargethispage{\baselineskip}

Let us now comment the subtleties related to the construction of the quasiparticle states. Recall that quantum state corresponding to the quasiparticle contains more than one additional particle excitation. As we have already mentioned, this is a statistical effect related to the fact that the background state is not an eigenstate of the number operator. It can be illustrated with  the following toy model. Let $\hat\rho = (|0\rangle\langle 0|+|2\rangle\langle 2|)/2$ be the initial state of a harmonic oscillator. When a ``quasiparticle'' is introduced into the system by the action of the creation operator, the state becomes $\hat\rho' = (|1\rangle\langle 1|+3|3\rangle\langle 3|)/4$. The expected number of particles in the initial and final states is 1 and 5/2 respectively, and therefore the particle number is increased in 3/2, despite the fact that each term of the state is increased with only one particle. Clearly, the reason for this  is linked to the Bose-Einstein statistics: the higher occupation states are more likely to be excited. A similar phenomenon happens if the incoherent mixture is replaced by a coherent superposition of two particle eigenstates. See Ref.~\cite{UnruhWald84} for an analysis of the same phenomenon in another context, with a discussion of possible energy non-conservation issues.

We have seen that the dispersion relations governing the particle dynamics can be also obtained from simpler mean-field-based methods (linear response theory or effective action), albeit with a slightly different interpretation. We believe that the effort put in the construction and analysis of the quasiparticle excitations was anyway useful for different reasons. First, the quasiparticle and mean field approaches provide the same answer to two different physical questions: in the former case one studies the dynamics of individual quasiparticles, and in the latter, the dynamics of the mean fields. Second, and related to this, both descriptions have different complementary regimes of validity, the former approach being suited for high-momentum perturbations and the latter being suited for low-momentum perturbations.  Finally, the route to the construction and analysis of the quasiparticle properties shed light on many intermediate results which are interesting by themselves. 

\index{Gaussian approximation}

A key element in our analysis has been the Gaussian approximation, which on the one hand has allowed manageable expressions, and on the other hand has provided physical interpretations for those expressions, without having to restore to any perturbative expansion in the coupling constant. The Gaussian truncation leads to the most basic description of the dynamics of the quasiparticles, the description in terms of a dispersion relation which we have investigated in this paper. Non-Gaussianities would appear in a more elaborate description of the quasiparticle dynamics.  The Gaussian approximation can be formally controlled with the large-$N$ expansion in the number of fields \cite{CopperMottola87,CooperEtAl94}. 

In this paper we have limited ourselves to those systems which can be correctly described in terms of a pole in the retarded propagator, or, in other words, in terms a dispersion relation. As we have commented, there are systems posessing massless excitations in which the analytic structure of the propagator differs, leading to non-exponential
 decay laws \cite{BlaizotIancu96,BlaizotIancu97,BoyanovskyDeVega98,BoyanovskyEtAl98} and to the so-called non-Fermi liquid behavior \cite{BoyanovskyDeVega01}. 
 
 \enlargethispage{\baselineskip}

A natural extension of the work in this paper would be to generalize the results for non-zero spin fields. The analysis of fermion fields should reveal the emergence of hole excitations as true quasiparticles, in contrast to the bosonic systems, in which we have seen that hole states, although can be constructed, do not have the appropriate quasiparticle properties. 
Another direction in which the work of this paper could be extended is the analysis of quasiparticle interactions.  In the same way as the K\"all\'en-Lehmann spectral representation can be adapted to non-vacuum situations, the Lehmann-Symanzik-Zimmerman formalism of vacuum field theory could in principle also be extended to non-vacuum situations. Some work has been already done in this direction in the context of thermofield dynamics \cite{NakawakiEtAl89}. 

The results of this paper can also be generalized  to study the propagation of interacting particles or quasiparticles in curved backgrounds, the propagation in curved backgrounds having many similarities with the propagation in a physical medium  \cite{ArteagaEtAl07a,Arteaga07a}. In Ref.~\cite{Arteaga08b} the main results of this paper are generalized to account for the adiabatic propagation of interacting particles in cosmological backgrounds. 

\subsection*{Acknowledgments}
	I am very grateful to Cristina Manuel, Guillem P\'erez-Nadal, Renaud Parentani, Albert Roura and Joan Soto for several helpful discussions and suggestions, and to Enric Verdaguer for his critical reading of the manuscript. This work is partially supported by the Research Projects MEC FPA2007-66665C02-02 and DURSI 2005SGR-00082.

\appendix
\chapter{The closed time path method} \label{app:CTP}

\index{Closed time path method|(}
\index{In-in method|see{closed time path method}}
\index{CTP|see{closed time path method}}

In this appendix we give a brief introduction to the closed time path (CTP) method (also called \emph{in-in} method, in contrast to the conventional \emph{in-out} method, or Keldysh-Schwinger method), stressing those aspects relevant for this paper. For further details we address the reader to Refs.~\cite{ChouEtAl85,CalzettaHu87,CamposHu98,CamposVerdaguer94,Weinberg05}.

Let us consider a free or an interacting scalar field $\phi$. The path-ordered generating functional $Z_{\mathcal C}[j]$ is defined
as
\begin{equation}
    Z_{\mathcal C}[j] = \Tr \left(\hat \rho T_{\mathcal C} \expp{i \int_{\mathcal C} \vd t \int
    \ud[3]{\vect
    x} \hat \phi(x) j(x)}  \right),
\end{equation}
where $\hat \phi(x)$ is the field operator in the Heisenberg
picture, ${\mathcal C}$ is a certain path in the complex $t$
plane, $T_{\mathcal C}$ means time ordering along this path and $j(x)$
is a classical external source.  By functional differentiation of
the generating functional with respect to $\phi$, path-ordered
correlation functions can be obtained:
\begin{equation}
	G_\mathcal C(x,x') = \Tr\big[ \hat\rho T_{\mathcal C} \hat\phi(x) \hat\phi(x')\big] =  - \left. \frac{ \delta^2  Z_{\mathcal C}}{\delta j(x) \delta j(x')}\right|_{j=0}.
\end{equation}
Introducing a complete basis of eigenstates of the field operator in the Heisenberg picture,
$	\hat\phi(t,\vect x) |\phi,t\rangle = \phi(t,\vect x) |\phi,t\rangle$, as a representation of the identity, the generating functional can be expressed as:
\begin{equation}  \label{PrePathGenFunct}
	Z_\mathcal C[j] = \int \widetilde{\mathrm d} \phi\, \widetilde{\mathrm d} \phi' \langle \phi,\ti|\hat\rho|\phi',\ti\rangle
	\langle \phi',\ti |T_{\mathcal C} \expp{i \int_{\mathcal C} \vd t \int
    \ud[3]{\vect
    x} \hat \phi(x) j(x)}  |\phi,\ti\rangle 
\end{equation}
The functional measures $\widetilde{\mathrm d} \phi$ and $\widetilde{\mathrm d} \phi'$ go over all field configurations of the fields at fixed initial time $t$. If the path $\mathcal C$ begins and ends at the same point $\ti$, then the transition element of the evolution operator can be computed via a path integral:
\begin{equation}\label{PathGenFunct}
	Z_\mathcal C[j] = \int \widetilde{\mathrm d} \phi\, \widetilde{\mathrm d} \phi' \langle \phi,\ti|\hat\rho|\phi',\ti\rangle
	\int_{\varphi(\ti,\vect x)=\phi(\vect x)}^{\varphi(\ti,\vect x)=\phi'(\vect x)} \mathcal D \varphi \expp{i \int_{\mathcal C} \vd t \int
    \ud[3]{\vect x}\{ L[\varphi] +  \varphi(x) j(x)\}},
\end{equation}
where $L[\phi]$ is the Lagrangian density of the scalar field.

\begin{figure}
    \centering
    \includegraphics{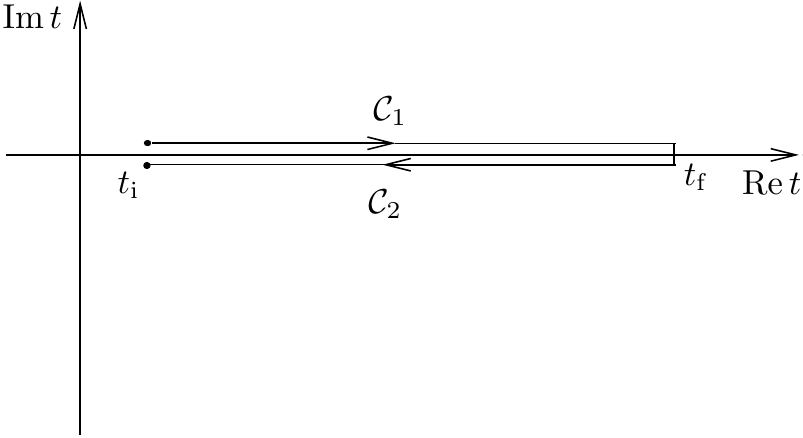}
    \caption{Integration path in the complex-time plane used in the
    CTP method. The forward and backward lines are infinitesimally close to the real axis.}
    \label{fig:CTP}
\end{figure}

Let us consider the time path shown in Fig.~\ref{fig:CTP}.  If we define $\varphi_{1,2}(t,\vect
x)=\varphi(t,\vect x)$ and $j_{1,2}(t,\vect x)=j(t,\vect x)$ for $t
\in {\mathcal C}_{1,2}$, then the generating functional can be reexpressed as:
\begin{equation}	 \label{CTPGenFunct}
\begin{split}
	Z[j_1,j_2] &= \int \widetilde{\mathrm d} \phi\, \widetilde{\mathrm d} \phi' \widetilde{\mathrm d} \phi'' \langle \phi,\ti|\hat\rho|\phi',\ti\rangle  \\
	&\qquad\times\int_{\varphi_1(\ti,\vect x)=\phi(\vect x)}^{\varphi_1(\tf,\vect x)=\phi''(\vect x)} \mathcal D \varphi_1 
	\expp{i \int \ud[4]{x}\{ L[\varphi_1] +  \varphi_1(x) j_1(x)\}}\\
	&\qquad\times\int_{\varphi_2(\ti,\vect x)=\phi'(\vect x)}^{\varphi_2(\tf,\vect x)=\phi''(\vect x)} \mathcal D \varphi_2 
	\expp{-i\int \ud[4]{x}\{ L[\varphi_2] +  \varphi_2(x) j_2(x)\}}.
\end{split}
\end{equation}
In the following it will prove useful to use a condensed notation where neither the boundary conditions of the path integral nor the integrals over the initial and final states are explicit. With this simplified notation the above equation becomes
\begin{equation}	
\begin{split}
	Z[j_1,j_2] &=  \int \mathcal D \phi_1 \mathcal D \phi_2\langle \phi,t|\hat\rho|\phi',t\rangle 
	\expp{i \int \ud[4]{x}\{ L[\varphi_1] - L[\varphi_2] +  \varphi_1(x) j_1(x)-  \varphi_2(x) j_2(x)\}}
\end{split}
\end{equation}
An operator representation is also possible:
\begin{equation}\label{ZCTPOper}
    Z[j_1,j_2] = \Tr 
	\left(\hat \rho \,  
	\widetilde T \expp{-i \int_\ti^\tf \vd t \int
    \ud[3]{\vect
    x} \hat \phi(x) j_2(x)} T \expp{i \int_\ti^\tf \vd t \int \ud[3]{\vect
    x} \hat \phi(x) j_1(x)} \right).
\end{equation}

\index{Propagator!Feynman}
\index{Propagator!Dyson}
\index{Propagator!Whightman}

By functionally differentiating the generating functional the different correlation functions can be obtained:
\begin{align}\label{CTPPropSet}
	G_{ab}(x,x')=  - \left. \frac{ \delta^2  Z}{\delta j^a(x) \delta j^b(x')}\right|_{j_a=j_b=0}.
\end{align}
 We recall that lowercase roman indices may acquire the values 1 and 2 are raised and lowered with the ``CTP metric''
$c_{ab}=\mathrm{diag}(1,-1)$. Higher order correlation functions can be obtained in a similar way.

At least formally, all expressions so far are valid either for interacting or free field theories. However, explicit results can only be obtained when the theory is free and the initial state is Gaussian. In this case the path integrals in \Eqref{CTPGenFunct} can be exactly performed, and one obtains:
\begin{equation}
	Z^{(0)}[j_1,j_2] = Z^{(0)}[0,0]  \expp{- \frac{1}{2} \int \ud[4]{x} \ud[4]{x'} j^a(x) G^{(0)}_{ab}(x,x') j^b(x')}. 
\end{equation}
where the propagators $G^{(0)}_{ab}$ verify $c^{ab}(-\partial_{\mu}\partial^\mu + m^2) G_{ab}^{(0)} =-i\delta^{(4)}(x-x')$.
As with the conventional in-out method, the  perturbative expansion can be organized in terms of Feynman diagrams. There are two kinds of vertices, type 1 and type 2, and four kinds of propagators linking the two vertices. The Feynman rules are those of standard scalar field theory, supplemented by the prescription of adding a minus sign for every type 2 vertex.

From the generating functional the connected generating functional $W[j,j']$ is defined
\begin{equation}
	W[j,j'] := - i \ln Z[j,j']
\end{equation}
Next we introduce the following objects:
\begin{equation}\label{CTPBarPhi}
	\varphi^a(x) = \frac{\delta W}{\delta j_a(x)},
\end{equation}
which must be understood as functionals of $j_1$ and $j_2$ even if this dependence is not explicit. If $j_1=j_2$ both $\phi_1$ and $\phi_2$ give the expectation value of the field under the presence of a classical source $j$. Finally, the effective action $\Gamma[\varphi_1,\varphi_2]$ is defined as the Legendre transform of the connected generating functional:
\begin{equation}
	\Gamma[\varphi_1,\varphi_2] :=  W[j_1,j_2] - \int \ud[4] x j_a(x) \varphi^a(x).
\end{equation}
In this equation $j_1$ and $j_2$ must be understood as functionals of $\varphi_1$ and $\varphi_2$, which can be obtained by inverting \Eqref{CTPBarPhi}.
By functionally differentiating the effective action with respect to $\varphi_1$ and setting $\varphi_2=\varphi_1$ the equation of motion for the expectation value of the scalar field is obtained:
\begin{equation}
	\left. \frac{\delta \Gamma}{\delta\varphi(x)} \right|_{\varphi_2=\varphi_1=0} = j(x).
\end{equation}
In contrast to the conventional \emph{in-out} treatment, the equations of motion obtained from the CTP generating functional are real and causal because they correspond to the dynamics of true expectation values \cite{Jordan86}.

\begin{figure}
    \centering
    \includegraphics{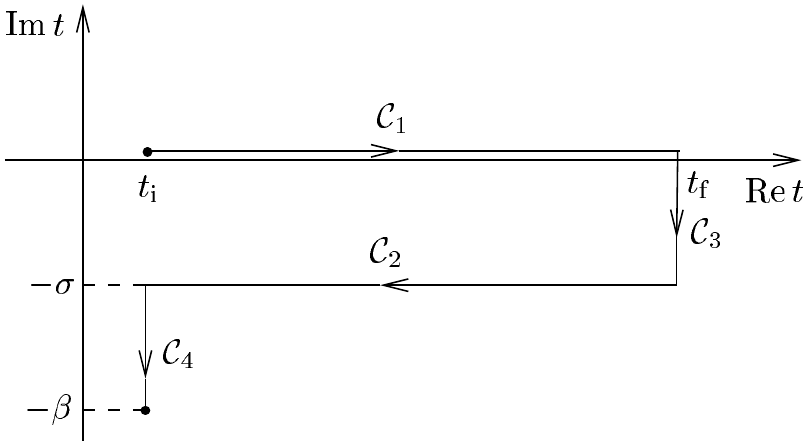}
    \caption{Integration contour in the complex-time plane used in the
    real-time approach to thermal field theory. The choice
    $\sigma=0^+$ makes the formalism analogous to the CTP method.}
    \label{fig:ThermalPath}
\end{figure}

Thermal field theory can be seen as a strict particular case of the CTP method, in which the state $\hat\rho$ happens to be 
\begin{equation}
     \hat\rho = \frac{\expp{- \beta\hat
    H}}{\Tr(\expp{- \beta\hat {H}})},
\end{equation}
where $\hat H$ is the Hamiltonian operator of the system. To apply the techniques presented in this appendix the only requirement is to compute the free thermal propagators. However, it proves convenient to make a slight a adaptation of the formalism and modify the complex time path in order to connect with the usual approaches to thermal field theory.
Noticing that the density matrix operator can be seen as the time translation operator in the complex plane, $\expp{-\beta\hat H} = U(t-i\beta,t)$, \Eqref{PrePathGenFunct} can be reexpressed as:
\begin{equation}  
\begin{split}
	Z_\mathcal C[j] = \frac{1}{\Tr(\expp{- \beta\hat {H}})} \int \widetilde{\mathrm d} \phi \langle \phi,\ti-i\beta|T_{\mathcal C} \expp{i \int_{\mathcal C} \vd t \int
    \ud[3]{\vect
    x} \hat \phi(x) j(x)}  |\phi,\ti\rangle ,
\end{split}
\end{equation}
where we have used the completeness relation $\int \widetilde{\mathrm d}\phi'|\phi',\ti\rangle\langle \phi',\ti|=1$. This way we have managed to incorporate the information on the state on the dynamical evolution. Now, if the path $\mathcal C$ starts at $\ti$ and ends at $\ti-i\beta$, a path-integral representation can be introduced:
\begin{equation}
\begin{split}
	Z_\mathcal C[j] &= \frac{1}{\Tr(\expp{- \beta\hat {H}})}
	\int \widetilde{\mathrm d} \phi 
	\int_{\varphi(\ti,\vect x)=\phi(\vect x)}^{\varphi(\ti,\vect x)=\phi(\vect x)} \mathcal D \varphi \expp{i \int_{\mathcal C} \vd t \int
    \ud[3]{\vect x}\{ L[\varphi] +  \varphi(x) j(x)\}},
\end{split}
\end{equation}
which can be rewritten in a more compact form as
\begin{equation}
\begin{split}
	Z_\mathcal C[j] &= N
	\int \mathcal D \varphi \expp{i \int_{\mathcal C} \vd t \int
    \ud[3]{\vect x}\{ L[\varphi] +  \varphi(x) j(x)\}},
\end{split}
\end{equation}
where $N$ is a normalization constant, and the boundary conditions $\phi(t_\mathrm
i,\vect x) = \phi(t_\mathrm i-i\beta,\vect x)$ are assumed.
Different elections for the path ${\mathcal C}$ lead to different approaches to
thermal field theory: a straight line from $t_\mathrm i$ to
$t_\mathrm i - i\beta$ leads to the imaginary-time formalism, and
the contour shown in Fig.~\ref{fig:ThermalPath} leads to the real-time
formalism. By choosing $\sigma= 0^+$ the real-time formalism is virtually identical to
the CTP method, since the  the path
along ${\mathcal C}_3$ and ${\mathcal C}_4$ can be neglected if we are interested in real-time correlation functions and the boundary
conditions of the path integral are properly taken into account.

\chapter{Gaussian states}\label{app:Gaussian}
\index{Gaussian state}

In the paper we make extensive use of the Gaussian states. In the following we shall give a brief description of the aspects relevant to us without entering into details. See \eg\ Refs.~\cite{WallsMilburn,GardinerZoller,Adam94} for a more complete introduction.
In this appendix $N$ will always represent the correct normalization constant, which can always be computed with the Gaussian integration formula if desired.

In general, Gaussian states are those whose density matrices have  a Gaussian form in a coordinate representation, or equivalently those whose Wigner functions have a Gaussian form. A generic Gaussian state with $\av{\hat p}=0$ and $\av{\hat q}=0$ (or equivalently with $\av{\hat a}=\av{\hat a^\dag}=0$) can be represented as:
\begin{equation}
	\hat\rho = N \exp{\big( -F \hat a^\dag \hat a + G \hat a \hat a + G^* \hat a^\dag \hat a^\dag \big)},
\end{equation}
where $F$ is real. In this paper we have exclusively used Gaussian states with zero mean (many times without making explicit mention). Gaussian stationary states are those which conmute with the Hamiltonian, $[\hat \rho,\hat H] = 0$. The stationary state of a harmonic oscillator with
$
	\hat H = 
	\Omega \big( \hat a^\dag \hat a + 1/2\big)
$ 
must be of the form $\hat\rho = N \exp{(-F  \hat a^\dag \hat a)}$, and therefore corresponds to a thermal state.

For a scalar field, decomposed in modes $\phi_\vect p$, the most general state for the two-mode system $\pm \vect p$ which is translation invariant, \ie, which conmutes with the momentum operator,\footnote{Do not confuse the  physical momentum operator $\hat{\vect p}$, with the canonical momentum operator $\hat\pi_\vect p$, conjugate of the field operator.}
\begin{equation}
	[\hat\rho_\text{s},\hat{\vect p}]=0, \qquad \hat{\vect p}= \vect p \big(\hat a^\dag_\vect p \hat a_\vect p  - \hat a^\dag_{-\vect p} \hat a_{-\vect p}\big).
\end{equation}
and which verifies $\av{\hat a_{\pm \vect p}}=\av{\hat a^\dag_{\pm \vect p}}=0$ is given by
\begin{equation}
	\hat\rho_\text{s} = N \exp{\big[
	-F \big( \hat a_\vect p^\dag \hat a_\vect p 
	+ \hat a_{-\vect p}^\dag \hat a_{-\vect p} \big)
	+ 2G \hat a_\vect p \hat a_{-\vect p} 
	+ 2G^* \hat a^\dag_\vect p \hat a^\dag_{-\vect p} \big]}.
\end{equation}
If we further impose stationarity with respect to the reduced Hamiltonian,
\begin{equation}
	[\hat\rho_\text{s},\hat H_\text{sys}]=0, \quad \hat H_\text{sys}= \hat\pi_\vect p \hat\pi_{-\vect p} + E^2_\vect p \hat\phi_\vect p \hat\phi_{-\vect p}  =E^2_\vect p \big(\hat a^\dag_\vect p \hat a_\vect p  + \hat a^\dag_{-\vect p} \hat a_{-\vect p} + 1\big),
\end{equation}
the most general Gaussian state corresponds to a factorized thermal state:
\begin{equation}
	\hat\rho_\text{s} = N \exp{\big[
	-F \big( \hat a_\vect p^\dag \hat a_\vect p 
	+ \hat a_{-\vect p}^\dag \hat a_{-\vect p} \big) \big]} =  N \exp{\big(
	-F  \hat a_\vect p^\dag \hat a_\vect p \big) } \exp{\big(
	-F  \hat a_{-\vect p}^\dag \hat a_{-\vect p} \big) }.
\end{equation}

A quantum mechanical or field theory system is Gaussian if its generating functional is Gaussian. For a single degree of freedom this means (assuming vanishing expectation values for the position operator)
\begin{equation} \label{AppGenF}
\begin{split}
	Z[j_{a}]&=\exp\bigg[-\frac{1}{2}\int \ud t \ud{t'} j^{a}(t)  G_{ab}(t,t')j^{b}(t) \bigg]
\end{split}
\end{equation}
Gaussian systems correspond either to Gaussian states following quadratic equations of motion (as in Ref.~\cite{Arteaga07b}), or alternatively to an approximation for general states following general equations of motion (as in this paper).

For Gaussian systems, according to the Wick theorem \cite{FetterWalecka,LeBellac}, the $n$-point correlation functions can be reduced to the two-point correlation functions. Instead of trying to give a general formulation of the Wick theorem let us simply show some particular applications.  
Whenever the system is Gaussian, any time-ordered four-point correlation function can be expressed in terms of two-point correlation functions (we  assume that the expectation value of the field operators vanishes):
\begin{equation}
\begin{split}
	\av{T\hat q(t_1) \hat q(t_2)\hat q(t_3)\hat q(t_4)}
	&= \av{T\hat q(t_1) \hat q(t_2)}\av{T\hat q(t_3)\hat q(t_4)} +
	\av{T\hat q(t_1) \hat q(t_3)}\av{T\hat q(t_2)\hat q(t_4)}
	\\ &\quad +
	\av{T\hat q(t_1) \hat q(t_4)}\av{T\hat q(t_2)\hat q(t_3)}
\end{split}
\end{equation}
If the correlation function is a mixture of time- and antitime-ordered expressions, the equivalent expression goes as follows
\begin{equation} \label{WickMixed}
\begin{split}
	\av{T\hat q(t_1) \hat q(t_2)\widetilde T\hat q(t_3)\hat q(t_4)}
	&= \av{T\hat q(t_1) \hat q(t_2)}\av{\widetilde T\hat q(t_3)\hat q(t_4)}  + \av{ \hat q(t_1)\hat q(t_3)}  \av{\hat q(t_2) \hat q(t_3)}\\
	&\quad  + \av{\hat q(t_1) \hat q(t_4)}\av{ \hat q(t_2)\hat q(t_3)} \\
\end{split}
\end{equation}
These expressions can be demonstrated by taking derivatives on the Gaussian generating functional \eqref{AppGenF}.

In field theory, if the state is translation-invariant, momentum conservation can simplify the application of the Wick theorem. For instance, if $\vect k \neq \vect q$,
\begin{equation}
\begin{split}
	\av{T\hat \phi_\vect k(t_1) \hat \phi_{-\vect q}(t_2)\hat \phi_\vect q(t_3)\hat \phi_{-\vect k}(t_4)}
	&= \av{T\hat \phi_\vect k(t_1) \hat \phi_{-\vect k}(t_4)}
	+ \av{T\hat \phi_\vect q(t_2) \hat \phi_{-\vect q}(t_3)}.
\end{split}
\end{equation}
The two-point correlators can be also expressed as a function of the creation and annihilation operator. If the state is stationary, only those terms having a creation and an annihilation operator survive:
\begin{equation}
\begin{split}
	\av{\hat \phi_\vect k (t_1) \hat \phi_{-\vect k}(t_2)} =  \frac{1}{2E_\vect p} &\Big(
	\av{\hat a^\dag_{-\vect k} (t_1) \hat a_{-\vect k} (t_2) } +
	\av{\hat a_{\vect k} (t_1) \hat a^\dag_\vect k  (t_2)}  \Big).
\end{split}
\end{equation}

\chapter{Contour integration of $I(t,t_0;\vect p)$} \label{app:Integral}

When computing the time evolution of the propagator, the following integral appears [see eqs.~\eqref{IntegralI} and \eqref{IntegralI2}]:
\begin{equation}
 I(t,t_0;\vect p) =  \frac{1}{2E_\vect p}\int \frac{\vd \omega}{2\pi} \expp{-i\omega(t-t_0)} \frac{ |\omega| (\omega+ E_\vect p)^2  \Gamma_\vect p \left[ n_\vect p + \theta(\omega)\right]  }{\left(-\omega^2+E_\vect p^2\right)^2 + \left( \omega \Gamma_\vect p\right)^2}
\end{equation}
Apparently, this integral cannot be evaluated with complex plane techniques since the integrand contains the factor and $|\omega|$, which is non-analytic with the usual prescription $|\omega| = \sqrt{\omega \omega^*}$, and the factor $\theta(\omega)$, which in principle is defined in the real axis. Let us do a more careful analysis. 

We begin by extending the problematic terms to the complex plane in the following way:
\[
	\theta(\omega) \to \theta(\Re \omega), \qquad
	|\omega| \to \omega \sign(\Re \omega).
\]
Those terms continue to be non-analytic, but only on a branch cut located in the imaginary axis. Therefore, with this prescription the integrand is analytic everywhere on the complex plane except on the branch cut and on the poles. Notice that the branch cut is rather special, since the function is continuous across the branch cut.

Second, notice that the following related contour integrals are well-defined for $t>t_0$
\begin{align*}
	I_3(t,t_0;\vect p) &=  \frac{1}{2E_\vect p}\oint\limits_{C_3} \frac{\vd \omega}{2\pi} \expp{-i\omega(t-t_0)} \frac{ -\omega (\omega+ E_\vect p)^2  \Gamma_\vect p  n_\vect p  }{\left(-\omega^2+E_\vect p^2\right)^2 + \left( \omega \Gamma_\vect p\right)^2}, \\
	I_4(t,t_0;\vect p) &=  \frac{1}{2E_\vect p}\oint\limits_{C_4}  \frac{\vd \omega}{2\pi} \expp{-i\omega(t-t_0)} \frac{ \omega (\omega+ E_\vect p)^2  \Gamma_\vect p ( n_\vect p +1)  }{\left(-\omega^2+E_\vect p^2\right)^2 + \left( \omega \Gamma_\vect p\right)^2},
\end{align*}
where $C_3$ and $C_4$ are the closed anticlockwise paths at the boundaries of the left lower and right lower quadrants respectively

Third, note that with the above prescription $I(t,t_0;\vect p) = I_3(t,t_0;\vect p) + I_4(t,t_0;\vect p)$, since the path at infinity does not contribute, and the contribution from the path in the imaginary axis cancels (because the function is continuous at the branch singularity). Therefore, in practice, the integral $I(t,t_0;\vect p)$ can be computed by residues as if there were no branch cut singularity.

Let us now compute the integrals $I_3(t,t_0;\vect p)$ and $I_4(t,t_0;\vect p)$. We start by $I_4(t,t_0;\vect p)$. There is a pole in the lower right quadrant at $\omega \approx E_\vect p - i \Gamma_\vect p/2$. Neglecting $\Gamma_\vect p$ in front of $E_\vect p$ we find
\begin{equation*}
	I_4(t,t_0;\vect p) \approx E_\vect p (1+n_\vect p) \expp{-iE_\vect p(t-t_0)} \expp{-\Gamma_\vect p/2}.
\end{equation*}
With respect to the integral $I_4(t,t_0;\vect p)$, the contribution from the pole in the lower left quadrant is very suppressed because of the factor $(\omega+E_\vect p)^2$ in the numerator. Therefore $I_3(t,t_0;\vect p) \ll I_4(t,t_0;\vect p)$. Given all this we obtain the final result:
\begin{equation}
	I(t,t_0;\vect p) \approx E_\vect p (1+n_\vect p) \expp{-iE_\vect p(t-t_0)} \expp{-\Gamma_\vect p/2}.
\end{equation}

Alternatively this integral can also be evaluated directly in the time domain \cite{Arteaga08b}.

\bibliography{articles,books}

\end{document}